\documentclass[aps,prl,reprint,superscriptaddress,showpacs,showkeys,ruledtabular]{revtex4-1}

\usepackage{graphicx}
\usepackage[usenames,dvipsnames,svgnames,table]{xcolor}
\definecolor{light-gray}{gray}{0.92}

\usepackage{color}

\begin{document}

\title{Wavelength-dispersive spectroscopy in the hard x-ray regime of a heavy highly-charged ion:
The 1s Lamb shift in hydrogen-like gold}

\author{T.\,Gassner}
%\thanks{thanks}
\affiliation{Helmholtz-Institut Jena, 07743 Jena, Germany}
\affiliation{GSI Helmholtzzentrum f\"ur Schwerionenforschung, 64291 Darmstadt, Germany}
\author{M.\,Trassinelli}
\email[corresponding author\,]{\textbf{martino.trassinelli@insp.jussieu.fr}}
\affiliation{Institut des NanoSciences de Paris, CNRS, Sorbonne Universit\'es - UPMC Univ Paris 06, 75005, Paris, France}
%\author{other}
\author{R.\,He\ss}
\affiliation{GSI Helmholtzzentrum f\"ur Schwerionenforschung, 64291 Darmstadt, Germany}
\author{U.\,Spillmann}
\affiliation{Helmholtz-Institut Jena, 07743 Jena, Germany}
\affiliation{GSI Helmholtzzentrum f\"ur Schwerionenforschung, 64291 Darmstadt, Germany}
\author{D.\,Bana\'{s}}
\affiliation{Institute of Physics, Jan Kochanowski University, PL-25406 Kielce, Poland}
\author{K.-H.\,Blumenhagen}
\affiliation{Helmholtz-Institut Jena, 07743 Jena, Germany}
\author{F.\,Bosch}\thanks{deceased 16.12.2016}
\affiliation{GSI Helmholtzzentrum f\"ur Schwerionenforschung, 64291 Darmstadt, Germany}
\author{C.\,Brandau}
\affiliation{GSI Helmholtzzentrum f\"ur Schwerionenforschung, 64291 Darmstadt, Germany}
\affiliation{I. Physikalisches Institut, Justus-Liebig-Universit\"at Gie{\ss}en, 35392 Gießen, Germany}
\author{W.\,Chen}
\affiliation{GSI Helmholtzzentrum f\"ur Schwerionenforschung, 64291 Darmstadt, Germany}
\author{Chr.\,Dimopoulou}
\affiliation{GSI Helmholtzzentrum f\"ur Schwerionenforschung, 64291 Darmstadt, Germany}
\author{E.\,F\"orster}
\affiliation{Helmholtz-Institut Jena, 07743 Jena, Germany}
\affiliation{Institut f\"ur Optik und Quantenelektronik, Friedrich-Schiller-Universit\"at, 07737 Jena, Germany}
\author{R.E.\,Grisenti}
\affiliation{GSI Helmholtzzentrum f\"ur Schwerionenforschung, 64291 Darmstadt, Germany}
\affiliation{Institut f\"ur Kernphysik, Goethe-Universit\"at, 60438 Frankfurt am Main, Germany}
\author{A.\,Gumberidze}
\affiliation{GSI Helmholtzzentrum f\"ur Schwerionenforschung, 64291 Darmstadt, Germany}
\affiliation{ExtreMe Matter Institute EMMI and Research Division, GSI Helmholtzzentrum f\"ur Schwerionenforschung, 64291 Darmstadt, Germany}
\author{S.\,Hagmann}
\affiliation{GSI Helmholtzzentrum f\"ur Schwerionenforschung, 64291 Darmstadt, Germany}
\affiliation{Institut f\"ur Kernphysik, Goethe-Universit\"at, 60438 Frankfurt am Main, Germany}
\author{P.-M.\,Hillenbrand}
\affiliation{GSI Helmholtzzentrum f\"ur Schwerionenforschung, 64291 Darmstadt, Germany}
\author{P.\,Indelicato}
\affiliation{Laboratoire Kastler Brossel, Sorbonne Universit\'es -  UPMC Univ Paris 06, ENS-PSL Research University, Coll\`ege de France, CNRS, 75005 Paris, France}
\author{P.\,Jagodzinski}
\affiliation{Department of Mathematics and Physics, Kielce University of Technology,  25-314 Kielce, Poland}
\author{T.\,K\"ampfer}
\affiliation{Helmholtz-Institut Jena, 07743 Jena, Germany}
\author{Chr.\,Kozhuharov}
\affiliation{GSI Helmholtzzentrum f\"ur Schwerionenforschung, 64291 Darmstadt, Germany}
\author{M.\,Lestinsky}
\affiliation{GSI Helmholtzzentrum f\"ur Schwerionenforschung, 64291 Darmstadt, Germany}
\author{D.\,Liesen}
\affiliation{GSI Helmholtzzentrum f\"ur Schwerionenforschung, 64291 Darmstadt, Germany}
\affiliation{Fakult\"at f\"ur Physik und Astronomie, Ruprecht-Karls-Universit\"at, 69117 Heidelberg, Germany}
\author{Yu.\,A.\,Litvinov}
\affiliation{GSI Helmholtzzentrum f\"ur Schwerionenforschung, 64291 Darmstadt, Germany}
\author{R.\,Loetzsch}
\affiliation{Helmholtz-Institut Jena, 07743 Jena, Germany}
\affiliation{Institut f\"ur Optik und Quantenelektronik, Friedrich-Schiller-Universit\"at, 07737 Jena, Germany}
\author{B.\,Manil}
\affiliation{Laboratoire de Physique des Lasers, CNRS, Universit\'{e} Paris 13, 93430 Villetaneuse, France}
\author{R.\,M\"artin}
\affiliation{Helmholtz-Institut Jena, 07743 Jena, Germany}
\author{F.\,Nolden}
\affiliation{GSI Helmholtzzentrum f\"ur Schwerionenforschung, 64291 Darmstadt, Germany}
\author{N.\,Petridis}
\affiliation{GSI Helmholtzzentrum f\"ur Schwerionenforschung, 64291 Darmstadt, Germany}
\affiliation{Institut f\"ur Kernphysik, Goethe-Universit\"at, 60438 Frankfurt am Main, Germany}
\author{M.\,S.\,Sanjari}
\affiliation{GSI Helmholtzzentrum f\"ur Schwerionenforschung, 64291 Darmstadt, Germany}
\author{K.S.\,Schulze}
\affiliation{Helmholtz-Institut Jena, 07743 Jena, Germany}
\affiliation{Institut f\"ur Optik und Quantenelektronik, Friedrich-Schiller-Universit\"at, 07737 Jena, Germany}
\author{M.\,Schwemlein}
\affiliation{Helmholtz-Institut Jena, 07743 Jena, Germany}
\author{A.\,Simionovici}
\affiliation{Institut des Sciences de la Terre, UGA, CNRS, CS 40700, 38058 Grenoble, France}
\author{M.\,Steck}
\affiliation{GSI Helmholtzzentrum f\"ur Schwerionenforschung, 64291 Darmstadt, Germany}
\author{Th.\,St\"ohlker}
\affiliation{Helmholtz-Institut Jena, 07743 Jena, Germany}
\affiliation{GSI Helmholtzzentrum f\"ur Schwerionenforschung, 64291 Darmstadt, Germany}
\affiliation{Institut f\"ur Optik und Quantenelektronik, Friedrich-Schiller-Universit\"at, 07737 Jena, Germany}
\author{C.\,I.\,Szabo}
\affiliation{Laboratoire Kastler Brossel, Sorbonne Universit\'es -  UPMC Univ Paris 06, ENS-PSL Research University, Coll\`ege de France, CNRS, 75005 Paris, France}
\affiliation{Theiss Research, 7411 Eads Ave, La Jolla, CA 92037, United States}
\author{S.\,Trotsenko}
\affiliation{Helmholtz-Institut Jena, 07743 Jena, Germany}
\author{I.\,Uschmann}
\affiliation{Helmholtz-Institut Jena, 07743 Jena, Germany}
\affiliation{Institut f\"ur Optik und Quantenelektronik, Friedrich-Schiller-Universit\"at, 07737 Jena, Germany}
\author{G.\,Weber}
\affiliation{Helmholtz-Institut Jena, 07743 Jena, Germany}
\author{O.\,Wehrhan}
\affiliation{Helmholtz-Institut Jena, 07743 Jena, Germany}
\author{N.\,Winckler}
\affiliation{GSI Helmholtzzentrum f\"ur Schwerionenforschung, 64291 Darmstadt, Germany}
\author{D.F.A.\,Winters}
\affiliation{GSI Helmholtzzentrum f\"ur Schwerionenforschung, 64291 Darmstadt, Germany}
\author{N.\,Winters}
\affiliation{GSI Helmholtzzentrum f\"ur Schwerionenforschung, 64291 Darmstadt, Germany}
\author{E.\,Ziegler}
\affiliation{European Synchrotron Radiation Facility, 38043 Grenoble, France}
\author{H.F.\,Beyer}
\affiliation{GSI Helmholtzzentrum f\"ur Schwerionenforschung, 64291 Darmstadt, Germany}

\date{\today}

\begin{abstract}
Accurate spectroscopy of highly charged high-Z ions in a storage
ring is demonstrated to be feasible by the use of specially
adapted crystal optics. The method has been applied for the
measurement of the 1s Lamb shift in hydrogen-like gold (Au$^{+78}$) in a storage ring through spectroscopy of the
Lyman x rays. This measurement represents the first result
obtained for a high-Z element using high-resolution
wavelength-dispersive spectroscopy in the hard x-ray regime,
paving the way for sensitivity to higher-order QED effects.
\end{abstract}


%\keywords{X-ray spectrometer, micro-strip detector, storage ring, high-Z ions, QED}

\maketitle

The theory of Quantum Electrodynamics (QED) has been tested for
light atoms with extraordinarily high accuracy
\cite{Fee1993,vanWijngaarden2000,Karshenboim2005,Parthey2011,CancioPastor2012,Notermans2014}.
Yet, in the recent years, measurements on muonic hydrogen
(combined with the state-of-the-art QED calculations), have produced
inconsistency with the results obtained from hydrogen spectroscopy
\cite{Pohl2010,Antognini2013}.
%This effect is commonly known as the ``proton radius puzzle'' and is still waiting for explanation.
The experimental verification of the QED predictions
is still significantly less precise in the domain of extreme field
strength as experienced by an electron bound to a nucleus with
high atomic number $Z$. In contrast to low-$Z$ ions, bound
state QED corrections are still a challenge for theory since they
have to be treated in all orders of $\alpha \, Z$. Here, a very recent measurement of the hyperfine splitting in hydrogen- and lithium-like Bismuth has hints at a large disagreement with the QED predictions \cite{Ullmann2017}.

The QED corrections to the electronic binding energy, made up by the self
energy and the vacuum polarization, are most important for the
inner shells of high-$Z$ systems since they approximately scale as
$Z^4/n^3$ \cite{Yerokhin2015}, where $n$ denotes the principal
quantum number. Hydrogen and hydrogen-like ions are the most
fundamental atomic systems where the QED effects can be calculated
with high accuracy, thus offering a possibility of stringent
experimental tests. From the experimental point of view it requires the preparation of
heavy hydrogen-like ions where notably the $1s$ Lamb shift can be accessed via x-ray spectroscopy of an $np \rightarrow 1s$
Lyman transition from which the calculated Dirac energy plus the
small QED contribution of the $np$ level are subtracted. Such
measurements have initially been performed at lower $Z$ where ion
intensities were sufficient for the use of high-resolution
techniques with low detection efficiency
\cite{Briand1983,Briand1983a,Briand1984,Richard1984,Beyer1985,Beyer1991}.
With the advent of heavy-ion accelerators and storage rings the
investigations could be extended to the highest nuclear charges up
to $Z=92$
\cite{Briand1990,Lupton1994,Stoehlker2000,Gumberidze2005}. However
the spectroscopy needed to be conducted with solid state Ge(i)
detectors ensuring a high detection efficiency, although they soon
faced their limits in spectral resolution. To circumvent the low resolving power of semiconductor detectors,
they were replaced by specially adapted crystal spectrometers, as
will be reported in this letter, and by calorimetric
low-temperature detectors yielding promising results in first
storage-ring experiments \cite{Kraft-Bermuth2017, Hengstler2015}.
In the present experiment a pair of crystal spectrometers was used
to measure the 1s Lamb shift of hydrogen-like gold accomplishing
for the first time both, high-$Z$ ions and high spectral
resolution. Envisioned for a long time, the measurements have
become feasible only recently because of the following
developments:  \textit{(i)} adapted and optimized
crystal-spectrometer optics, \textit{(ii)} specially developed
two-dimensionally position sensitive Ge(i) detectors for hard x
rays with both energy and time resolution and \textit{(iii)} a
substantial increase of the ion-beam intensity in the Experimental Storage Ring (ESR) \cite{Beyer2004,Chatterjee2006}.

\begin{figure}[t]
\centering
\includegraphics*[width=0.9\columnwidth]{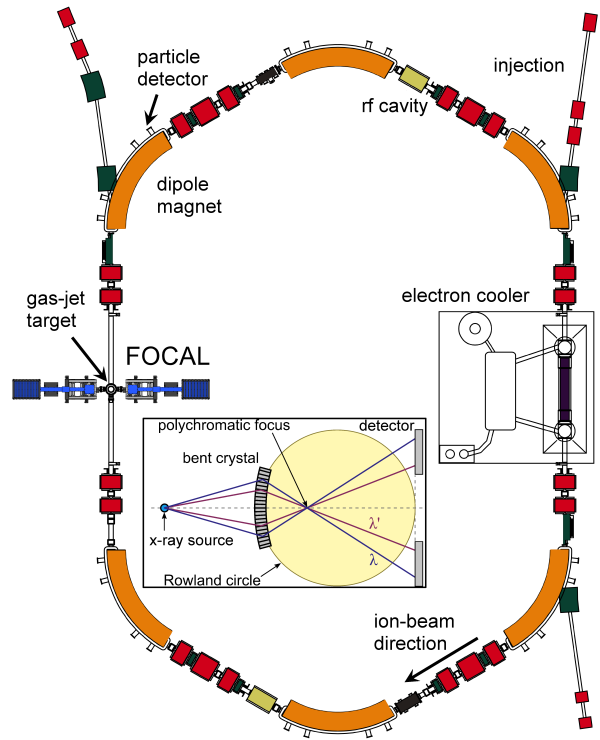}
\vspace*{-1mm} \caption{Schematic view of the ESR storage ring and the location of the FOCAL spectrometer at the gas-jet target. 
In the upper left dipole magnet a particle detector recording down-charged ions is used to apply a coincidence condition on the
x-ray spectra. The inset shows a schematic view of the FOCAL crystal-optics layout.}
\label{fig:ESR}
\end{figure}

The experiment was performed at the accelerator and storage-ring facility of the GSI Helmholtz Centre for Heavy Ion Research in Darmstadt, Germany \cite{Blasche1989}. Up to 10$^8$ of fully ionized gold atoms (Au$^{+79}$) per pulse with an initial kinetic energy of about 300\,MeV per nucleon were injected into the ESR experimental storage ring (see Fig.~\ref{fig:ESR}). Here, they were stored, cooled, and decelerated to a final velocity of $\beta = v_{ion}/c = 0.471\,36(10)$. The relative momentum spread ($\Delta p/p$) of the cooled ion beam is typically in the range of $10^{-4} -10^{-5}$. The cooled Au$^{+79}$ ions were then brought into interaction with the ESR internal gas target in the form of a supersonic gas jet overlapping with the circulating ion beam. A typical gas area density of $\sim 10^{12}\, \text{atoms/cm}^2$ guaranteed single collision conditions and a reasonably long ion beam storage times of several tens of seconds. In the present experiments, argon and krypton have been used as target gases. During each collision one ion has a chance to capture an electron from the target atom into an excited state, which then decays either directly or in a very rapid cascade to the $1s$ ground state of the newly formed hydrogen-like ion. About 1/3 of all down-charged ions decay (among other transitions in the cascade) via the Lyman-$\alpha_1$ ($2p_{3/2} \rightarrow 1s$) transition, the accurate spectroscopy of which is the main goal of the present experiment.

The Lyman-$\alpha_1$ transition wavelength is measured by two twin spectrometers operated in the focussing compensated Laue (FOCAL) geometry \cite{Beyer2009, Beyer2015}.
This type of spectrometer is well suited to find the right compromise between superior spectral resolving power and sufficient detection efficiency in the situation of very limited source strength and the presence of strong Doppler effects. Since the radiation source moves with relativistic velocity relative to the resting detector assembly (the laboratory frame) the velocity and observation-angle dependent Doppler effect has to be taken into account. The velocity of the ion beam is set by the electron cooler, however it seems unfeasible to aim for a determination of the actual observation angle with comparable accuracy. 
%Due to this reason the dedicated FOCAL twin-crystal-spectrometer layout was developed \cite{Beyer2009, Beyer2015}, where the
The two identical crystal spectrometer arms are aligned perpendicular with respect to the ion beam at both sides of the interaction chamber on one common line of sight. Both spectrometers 
%(called FOCAL 1 and 2) 
are used to measure the Lyman-$\alpha_1$ transition independently of each other leading to two distinct results for the wavelength $\lambda_{1,2}$. In this special geometry the observation-angle dependency of the Doppler equation cancels out and the rest-frame transition wavelength $\lambda_0$ can be derived via
\begin{equation}
\lambda_{\text{1}}+\lambda_{\text{2}} = 2 \, \gamma \, \lambda_{\text{0}},
\label{eqn:doppler_cancelation}
\end{equation}
with the velocity dependent Lorentz factor $\gamma$.

The wavelengths $\lambda_{1,2}$ are measured with respect to a calibration line from an isotope enriched $^{169}$Yb source. The strong and well known $63\,120.44(4)\,\text{eV}$ $\gamma$ transition \cite{Be2004} was selected as the main calibration line. The ion-beam velocity has been chosen such ($\beta=0.471\,36(10)$), that  the Doppler-shifted lab-frame energy of the Lyman-$\alpha_1$ transition approximately coincides with this calibration energy thus avoiding systematic uncertainties due to large extrapolations.
The wavelength comparison is made with respect to the dispersion plane defined by the crystals and detectors of the twin spectrometers.

The actual crystal-optics layout of each FOCAL spectrometer arm is shown in the inset of Fig.~\ref{fig:ESR}. The emitted
x-ray radiation is Bragg diffracted by the cylindrically bent silicon single
crystal, with a bending radius of $2\,\text{m}$ \cite{Beyer2009}.
The diffracted x rays cross the polychromatic focus and are recorded in one of the
position sensitive x-ray detectors. Due to the curvature of the crystal, the spatially wide x-ray radiation is focused to
a narrow line at the edge of the Rowland circle whose diameter is
equal to the crystal bending radius.
The intensity of x rays emitted from the Au$^{78+}$ reaction products is too faint in order to allow the usage of a conventional crystal spectrometer geometry.
For this purpose an {\em asymmetric} crystal cut has been applied with an angle deviation of $\chi=2^\circ$ from the symmetric Laue case, where the reflecting lattice planes are orientated perpendicular with respect to the principal crystal faces. This asymmetric cut leads to a broadening 
%of the Rocking curve and hence to an increased integrated reflectivity \cite{Beyer2009}. This way the efficiency could be enhanced by more than a factor of 20. 
of the crystal reflection curve, thereby enhancing the efficiency by more than a factor of 20~\cite{Beyer2009}.
The bent crystal is rotated by the angle $\chi$ to correct for the asymmetric cut, leading to symmetric but mirrored reflections above and below the optical axis.

The position sensitive x-ray detectors are located close to the Rowland circle
to make use of this focusing effect. Each 
%FOCAL 
spectrometer arm is equipped with one germanium microstrip detector consisting of an 11-mm-thick germanium single crystal with both anode and cathode
segmented into many strips \cite{Spillmann2008}. The cathode is divided into 128
56-mm-wide and 250-$\mu$m-high strips, whereas the anode is
segmented into 48 1.2-mm-wide and 32-mm-high strips. The strips on the front
and on the back side are oriented perpendicularly with respect to
each other allowing a two dimensional position reconstruction if
front and back side strips are combined for events with the same
measured energy. The narrow
strips on the front are orientated perpendicularly to the
dispersive direction of the spectrometer allowing for a more
accurate position determination.

Both spectrometers are passively shielded by 15\,mm thick lead plates and several thick tungsten diaphragms along the ray path to ensure that the majority of the detected photons stem from a diffractive process from the crystal. Additional background suppression was achieved by active shielding making use of the fact that the down-charged ions follow a different trajectory in the bending dipole magnets of the ESR, where they were recorded by a particle detector with efficiency close to 100 \%.  X-ray events in the germanium detectors have been taken into account only if a singly down-charged ion has been coincidently detected in the particle detector.
\begin{figure}[t]
\centering
\includegraphics*[width=0.85\columnwidth]{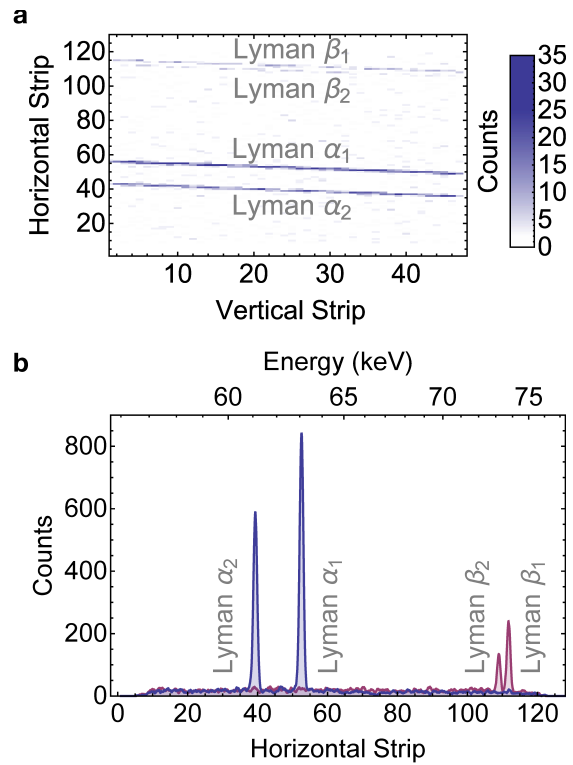}
\vspace*{-1mm}
\caption{(a) Coincident x-ray spectrum as recorded by one of the FOCAL spectrometers. (b) Spectrum of the Lyman-$\alpha$ and -$\beta$ doublets of Au$^{78+}$ obtained by a projection of the respective two-dimensional intensity distribution shown in (a).}
\label{fig:Lyman}
\end{figure}

Figure 2(a) shows the Lyman spectrum of H-like gold as measured by one of the two spectrometers by applying appropriate energy and time coincidence conditions to the data. In this way almost background free lines are revealed as can be seen in the figure. In three weeks of almost interruption free data taking about 1500 Lyman-$\alpha_1$ photons per spectrometer arm could be collected. In addition to the Lyman-$\alpha$, also the spatially resolved Lyman-$\beta$ transitions could be recorded, clearly evidencing the high resolving power of FOCAL. The slight tilt of the lines over several horizontal strips is caused by an effect called Doppler slating. Due to the spatial extent of the 2D detector a certain observation angle interval is covered, leading to higher Doppler shifted (laboratory frame) transition energies in forward angles relative to the backward direction. Figure \ref{fig:Lyman}(b) shows the spectrum obtained by projection of the 2D image in Fig.~\ref{fig:Lyman}(a) according to the tilt angle.

Since the spectrometer is operated as a wavelength comparator only the relative distance $\Delta z_\mathrm{d}$ between the main $^{169}$Yb-$\gamma$ calibration line and the Lyman-$\alpha_1$ line matters. This distance was determined by fitting a 2D model function to the original (not projected) measurement data for the Lyman and the $^{169}$Yb calibration data. The fitting results can be found in table \ref{tab:fitting_results}. Here the minus sign indicates that the measured laboratory frame energy lies below the $^{169}$Yb-$\gamma$ line energy. Possible model dependencies and details of the line shape have also been addressed \cite{Beyer2015} by applying various fitting procedures resulting in 
%only minor uncertainties in the line position which are subsumed under the statistical uncertainties.
negligible uncertainties.

\begin{table} [b]
\caption{Line Spacing between the main $^{169}$Yb-$\gamma$ calibration line and the Lyman-$\alpha_1$ transition, and the measured spectrometer dispersion.}
\begin{center}
\vspace*{-6pt}
\begingroup\setlength{\fboxsep}{0pt}
 \begin{ruledtabular}
\begin{tabular*}{\columnwidth}{lcc}
 & $\Delta z_\mathrm{d}$ ($\mu$m) & Spectrometer Dispersion  \\
\hline
FOCAL 1 \rule{0pt}{10pt} & $-35.2(5.1)$ & $1.905\,29(53)\times 10^{-10}$ \\
FOCAL 2 & $-51.8(3.6)$ & $1.909\,74(52)\times 10^{-10}$ \\
\end{tabular*}
\end{ruledtabular}
\endgroup
\end{center}
\label{tab:fitting_results}
\end{table}

Besides the line spacing also the spectrometers dispersion $D$ for both assemblies has to be measured. This was done by fitting in addition to the main $^{169}$Yb-$\gamma$ calibration line, the thulium K$\beta_{1,3}$ transitions, which are present in the calibration source spectra. The results are listed in table \ref{tab:fitting_results}.

By using Eq.~(\ref{eqn:doppler_cancelation}), with thus obtain a preliminary Lyman-$\alpha_1$ transition energy of $E_{Ly-\alpha_1}^\mathrm{prel.}=71\,539.8(2.2)\,\text{eV}$, which does not include any systematic effects so far.
%
%In addition to the already mentioned statistical uncertainties a number of systematic effects has to be considered, which do not only increase the total uncertainty but may also shift the final value of the Lyman-$\alpha_1$ transition energy. In the following a survey of the performed investigations is given, and a compilation of the corresponding systematic effects and uncertainties are given in table \ref{tab:systematic_effects}.
%This value does not include any systematic effects, which 
The systematic effects do not only increase the total uncertainty but may also shift the final value of the Lyman-$\alpha_1$ transition energy. All possible contributions are discussed below and are listed in table \ref{tab:systematic_effects} with the corresponding estimated uncertainties.

The first systematic effect was the temporal drift of the assembly during the three weeks of beam time. The drifts were monitored by the $^{169}$Yb calibrations which were done every six hours. In total it was less than 100\,$\mu$m for both spectrometers. With the help of the numerous calibrations the effect could be minimized, adding $\pm$2.8\,eV to the total uncertainty.

If the moving x-ray source is shifted along the common line of sight between the two spectrometer arms, the FOCAL geometry corrects for that effect. However, if the source position
(\textit{i.e.} ion-beam--gas-target intersection region) is shifted out of that line (\textit{i.e.} along the beam direction) this misalignment can not be corrected leading to a systematic deviation.
For the actual position measurement of the gas-jet target a dedicated auxiliary experiment was performed in the aftermath of the 
%FOCAL 
beam time \cite{Gassner2015} and
 the position of the gas-jet target was measured with an uncertainty of $\pm0.30$\,mm revealing an offset of 0.25\,mm in the ion beam direction. The corrected gas-jet position represents our present best guess. However, because of the long time passed between the main and the auxiliary experiment, we need to increase the position uncertainty to $\pm1$\,mm in order to account for possible long-time changes (due to mechanical manipulations, venting and pumping, \textit{etc.}). Fluctuations of this magnitude have previously been observed when checking the optical alignment of the gas-jet nozzles or when measuring the maximum overlap of the ion beam with the gas-jet. For the Lyman-$\alpha_1$ transition energy it means a correction of 3.2\,eV with an associated uncertainty of $\pm13$\,eV.

Also the uncertainty in the ion-beam velocity has to be considered, which is mainly caused  by an insufficiently accurate calibration of the high-voltage terminal of the electron cooler. 
Another correction to be added is due to the space charge of the electron beam. Details concerning the evaluation of these corrections and the associated uncertainties can be found in \cite{Lochmann2013,Lochmann2014,Brandau2000}.
The influence on the total uncertainty of this systematic effect is $\pm$4.3\,eV.

The last and strongest influence on the final value is given by
the actual position of the germanium detector crystal inside the
housing of the position sensitive x-ray detector. For its
measurement, a dedicated beam time at the European Synchrotron
Radiation Facility (ESRF) in Grenoble, France, has been conducted
where an intense and narrow x-ray beam can be provided. The
position sensitive detector was mounted on a movable table
directly facing the x-ray beam. In small steps the detector was
moved and the count rate of the strips as a function of the
detector position was recorded. The
location of the germanium crystal was thus measured relative to the outer fiducial
marks, which were also used during the original experimental assembly. The findings from the ESRF
measurement lead to a systematic energy decrease of 11.6\,eV with
an uncertainty of $\pm$5.1\,eV.

\begin{table} [b]
\caption{Individual contributions to the total Lyman-$\alpha_1$ transition energy.}
\begin{center}
\vspace*{-6pt}
\begingroup\setlength{\fboxsep}{0pt}
 \begin{ruledtabular}
\begin{tabular*}{\columnwidth}{lr}
Contribution & Value (eV)\\
\hline
Preliminary Transition Energy \rule{0pt}{10pt} & $71\,539.8(2.2)$\\
Temporal Drift & --(2.8) \\
Gas-Target Position & +3.2(13.0) \\
Ion-Beam Velocity & --(4.3)\\
Detector-Crystal Position & $-$11.6(5.1)\\
\hline
\textbf{Total} \rule{0pt}{10pt} & \textbf{71\,531.5(15.0)}\\
\end{tabular*}
\end{ruledtabular}
\endgroup
\end{center}
\label{tab:systematic_effects}
\end{table}

Our final experimental value for the Lyman-$\alpha_1$ transition energy including all statistical and systematic uncertainties (added quadratically) is given by $E_{Ly-\alpha_1}^\mathrm{exp.}=71\,531.5(15.0)\,\text{eV}$.

\begin{table} [b]
\caption{The $1s$ Lamb shift of Au$^{78+}$ in eV.}
\begin{center}
\vspace*{-6pt}
\begingroup\setlength{\fboxsep}{0pt}
\begin{ruledtabular}
\begin{tabular}{lr}
Present Experiment & 244.1(15.0)\\
Beyer \textit{et al.} 1995  \cite{Beyer1995} & 202.3(7.9)\\
Kraft-Bermuth \textit{et al.}  2016 \cite{Kraft-Bermuth2017} & 211(42) \\[5pt]
Theory, Yerokhin and Shabaev 2015 \cite{Yerokhin2015} & 205.2(2)\\
\end{tabular}
\end{ruledtabular}
\endgroup
\end{center}
\label{tab:lambshift}
\end{table}

The experimental value for the $1s$ Lamb shift is obtained by
subtracting the FOCAL value for the Lyman-$\alpha_1$ transition
energy from the theoretical value for the $2p_{3/2}$ binding
energy, which is sufficiently well known \cite{Yerokhin2015}. The
difference between this value and the Dirac value for the $1s$
binding energy yields the $1s$ Lamb shift. With the theoretical
value for the $2p_{3/2}$ binding energy
$E_{2p_{3/2}}^\mathrm{theo.}=-21\,684.201(5)\,\text{eV}$ one
obtains $\Delta E_{1s}^\mathrm{exp.}=244.1(15.0)\,\text{eV}$. In
Table \ref{tab:lambshift} our result is compared to the
experimental value obtained with a Ge(i) detector in an early
experiment at the ESR electron cooler  \cite{Beyer1995} and to the
experimental result reported for the calorimetric low-temperature
detectors which was gained in the same beam time
\cite{Kraft-Bermuth2017} as our present experiment. In the last
entry of the Table the theoretical value of Yerokhin and Shabaev
\cite{Yerokhin2015} is given. Our present value of the Lamb shift
is higher than the theoretical value and the other experimental
results by about 2.5 standard deviations of the estimated
experimental uncertainty.

It is difficult at this stage to unambiguously pinpoint the reason behind this deviation. Without going into details of the other results which would be beyond the scope of this article, one can say that each of the measurements has been performed with different techniques, \textit{i.e.} semiconductor detector at the electron cooler \cite{Beyer1995}, microcalorimeter at the gas jet target \cite{Kraft-Bermuth2017} and thus are prone to different systematic effects. It is important to emphasize that even though we have performed very thorough and extensive studies of the various possible systematic effects, since this is the first measurement of its kind at the storage ring, potentially underestimated or unknown systematic effects can not be fully excluded. Therefore more measurements are required in order to clarify this issue.

In conclusion we performed a first measurement of the ground-state
Lamb shift in a heavy H-like ion (Au$^{78+}$) using a high
resolution crystal spectrometer in combination with a fast and dim
source of hard x rays as present at a heavy-ion storage ring. The
energy resolution corresponding to about 60\,eV FWHM at 63\,keV
photon energy \cite{Beyer2015} surpasses the best semiconductor
detectors by almost one order of magnitude. The achieved
statistical uncertainty of 2.2\,eV is groundbreaking for a crystal
spectrometer operated in the region of hard x rays of H-like
high-$Z$ ions. Since storage rings are currently the only
facilities routinely delivering high-Z hydrogen-like ions in large
quantities, this measurement represents a very important milestone
towards the challenging goal of achieving a sensitivity to
higher-order QED effects as it is planned at the FAIR facility
\cite{Stoehlker2015}. In a future run, particular effort has to be
put into avoiding or reducing systematic uncertainties. The
ion-beam velocity can already be determined with a much higher
accuracy using a high-voltage divider from the
Physikalisch-Technische Bundesanstalt (PTB) in the electron-cooler
terminal, which will establish an absolutely calibrated velocity
standard \cite{Hallstrom2014}. With a slightly modified assembly it will also be
possible to measure the gas-target position relative to the
detector-crystal position \textit{in situ}, which will almost
entirely eliminate these systematic uncertainties avoiding
supplementary experiments alltogether.

Furthermore, we would like to emphasize that this apparatus can also
be applied for precision spectroscopy of heaviest helium-like ions
(as well as other few-electron systems) which, taking into account
the unprecedented resolution, would allow for resolving all the
relevant fine structure levels for the first time. This is
especially interesting in the light of the recent controversy with
the comparison between the experimental and theoretical results
for helium-like ions
\cite{Trassinelli2009,Chantler2012,Amaro2012,Rudolph2013,Kubicek2014,Payne2014,Epp2015,Beiersdorfer2015}.

\begin{acknowledgments}
Laboratoire Kastler Brossel (LKB) is ``Unit\'{e} Mixte de Recherche de Sorbonne University-UPMC, de ENS-PSL Research University, du Collège de France et du CNRS n$^{\circ}$ 8552''.
Institut des NanoSciences de Paris (INSP) is ``Unit\'{e} Mixte de Recherche de Sorbonne University-UPMC et du CNRS n$^{\circ}$ 7588''.
This work has been partially supported by:  the European Community FP7 - Capacities, contract ENSAR n$^{\circ}$ 262010, the Allianz Program of the Helmholtz Association contract n$^{\circ}$ EMMI HA-216 ``Extremes of Density and Temperature: Cosmic Matter in the Laboratory, the Helmholtz-CAS Joint Research Group HCJRG-108 and by the German Ministry of Education and Research (BMBF) under contract 05P15RGFAA.
\end{acknowledgments}

\bibliography{Gold-MT2}

\begin{thebibliography}{44}%
\makeatletter
\providecommand \@ifxundefined [1]{%
 \@ifx{#1\undefined}
}%
\providecommand \@ifnum [1]{%
 \ifnum #1\expandafter \@firstoftwo
 \else \expandafter \@secondoftwo
 \fi
}%
\providecommand \@ifx [1]{%
 \ifx #1\expandafter \@firstoftwo
 \else \expandafter \@secondoftwo
 \fi
}%
\providecommand \natexlab [1]{#1}%
\providecommand \enquote  [1]{``#1''}%
\providecommand \bibnamefont  [1]{#1}%
\providecommand \bibfnamefont [1]{#1}%
\providecommand \citenamefont [1]{#1}%
\providecommand \href@noop [0]{\@secondoftwo}%
\providecommand \href [0]{\begingroup \@sanitize@url \@href}%
\providecommand \@href[1]{\@@startlink{#1}\@@href}%
\providecommand \@@href[1]{\endgroup#1\@@endlink}%
\providecommand \@sanitize@url [0]{\catcode `\\12\catcode `\$12\catcode
  `\&12\catcode `\#12\catcode `\^12\catcode `\_12\catcode `\%12\relax}%
\providecommand \@@startlink[1]{}%
\providecommand \@@endlink[0]{}%
\providecommand \url  [0]{\begingroup\@sanitize@url \@url }%
\providecommand \@url [1]{\endgroup\@href {#1}{\urlprefix }}%
\providecommand \urlprefix  [0]{URL }%
\providecommand \Eprint [0]{\href }%
\providecommand \doibase [0]{http://dx.doi.org/}%
\providecommand \selectlanguage [0]{\@gobble}%
\providecommand \bibinfo  [0]{\@secondoftwo}%
\providecommand \bibfield  [0]{\@secondoftwo}%
\providecommand \translation [1]{[#1]}%
\providecommand \BibitemOpen [0]{}%
\providecommand \bibitemStop [0]{}%
\providecommand \bibitemNoStop [0]{.\EOS\space}%
\providecommand \EOS [0]{\spacefactor3000\relax}%
\providecommand \BibitemShut  [1]{\csname bibitem#1\endcsname}%
\let\auto@bib@innerbib\@empty
%</preamble>
\bibitem [{\citenamefont {Fee}\ \emph {et~al.}(1993)\citenamefont {Fee},
  \citenamefont {Chu}, \citenamefont {Mills}, \citenamefont {Chichester},
  \citenamefont {Zuckerman}, \citenamefont {Shaw},\ and\ \citenamefont
  {Danzmann}}]{Fee1993}%
  \BibitemOpen
  \bibfield  {author} {\bibinfo {author} {\bibfnamefont {M.~S.}\ \bibnamefont
  {Fee}}, \bibinfo {author} {\bibfnamefont {S.}~\bibnamefont {Chu}}, \bibinfo
  {author} {\bibfnamefont {A.~P.}\ \bibnamefont {Mills}}, \bibinfo {author}
  {\bibfnamefont {R.~J.}\ \bibnamefont {Chichester}}, \bibinfo {author}
  {\bibfnamefont {D.~M.}\ \bibnamefont {Zuckerman}}, \bibinfo {author}
  {\bibfnamefont {E.~D.}\ \bibnamefont {Shaw}}, \ and\ \bibinfo {author}
  {\bibfnamefont {K.}~\bibnamefont {Danzmann}},\ }\href@noop {} {\bibfield
  {journal} {\bibinfo  {journal} {Phys. Rev. A}\ }\textbf {\bibinfo {volume}
  {48}},\ \bibinfo {pages} {192} (\bibinfo {year} {1993})}\BibitemShut
  {NoStop}%
\bibitem [{\citenamefont {van Wijngaarden}\ \emph {et~al.}(2000)\citenamefont
  {van Wijngaarden}, \citenamefont {Holuj},\ and\ \citenamefont
  {Drake}}]{vanWijngaarden2000}%
  \BibitemOpen
  \bibfield  {author} {\bibinfo {author} {\bibfnamefont {A.}~\bibnamefont {van
  Wijngaarden}}, \bibinfo {author} {\bibfnamefont {F.}~\bibnamefont {Holuj}}, \
  and\ \bibinfo {author} {\bibfnamefont {G.~W.~F.}\ \bibnamefont {Drake}},\
  }\href@noop {} {\bibfield  {journal} {\bibinfo  {journal} {Phys. Rev. A}\
  }\textbf {\bibinfo {volume} {63}},\ \bibinfo {pages} {012505} (\bibinfo
  {year} {2000})}\BibitemShut {NoStop}%
\bibitem [{\citenamefont {Karshenboim}(2005)}]{Karshenboim2005}%
  \BibitemOpen
  \bibfield  {author} {\bibinfo {author} {\bibfnamefont {S.~G.}\ \bibnamefont
  {Karshenboim}},\ }\href@noop {} {\bibfield  {journal} {\bibinfo  {journal}
  {Phys. Rep.}\ }\textbf {\bibinfo {volume} {422}},\ \bibinfo {pages} {1}
  (\bibinfo {year} {2005})}\BibitemShut {NoStop}%
\bibitem [{\citenamefont {Parthey}\ \emph {et~al.}(2011)\citenamefont
  {Parthey}, \citenamefont {Matveev}, \citenamefont {Alnis}, \citenamefont
  {Bernhardt}, \citenamefont {Beyer}, \citenamefont {Holzwarth}, \citenamefont
  {Maistrou}, \citenamefont {Pohl}, \citenamefont {Predehl}, \citenamefont
  {Udem}, \citenamefont {Wilken}, \citenamefont {Kolachevsky}, \citenamefont
  {Abgrall}, \citenamefont {Rovera}, \citenamefont {Salomon}, \citenamefont
  {Laurent},\ and\ \citenamefont {H\"ansch}}]{Parthey2011}%
  \BibitemOpen
  \bibfield  {author} {\bibinfo {author} {\bibfnamefont {C.~G.}\ \bibnamefont
  {Parthey}}, \bibinfo {author} {\bibfnamefont {A.}~\bibnamefont {Matveev}},
  \bibinfo {author} {\bibfnamefont {J.}~\bibnamefont {Alnis}}, \bibinfo
  {author} {\bibfnamefont {B.}~\bibnamefont {Bernhardt}}, \bibinfo {author}
  {\bibfnamefont {A.}~\bibnamefont {Beyer}}, \bibinfo {author} {\bibfnamefont
  {R.}~\bibnamefont {Holzwarth}}, \bibinfo {author} {\bibfnamefont
  {A.}~\bibnamefont {Maistrou}}, \bibinfo {author} {\bibfnamefont
  {R.}~\bibnamefont {Pohl}}, \bibinfo {author} {\bibfnamefont {K.}~\bibnamefont
  {Predehl}}, \bibinfo {author} {\bibfnamefont {T.}~\bibnamefont {Udem}},
  \bibinfo {author} {\bibfnamefont {T.}~\bibnamefont {Wilken}}, \bibinfo
  {author} {\bibfnamefont {N.}~\bibnamefont {Kolachevsky}}, \bibinfo {author}
  {\bibfnamefont {M.}~\bibnamefont {Abgrall}}, \bibinfo {author} {\bibfnamefont
  {D.}~\bibnamefont {Rovera}}, \bibinfo {author} {\bibfnamefont
  {C.}~\bibnamefont {Salomon}}, \bibinfo {author} {\bibfnamefont
  {P.}~\bibnamefont {Laurent}}, \ and\ \bibinfo {author} {\bibfnamefont
  {T.~W.}\ \bibnamefont {H\"ansch}},\ }\href@noop {} {\bibfield  {journal}
  {\bibinfo  {journal} {Phys. Rev. Lett.}\ }\textbf {\bibinfo {volume} {107}},\
  \bibinfo {pages} {203001} (\bibinfo {year} {2011})}\BibitemShut {NoStop}%
\bibitem [{\citenamefont {Cancio~Pastor}\ \emph {et~al.}(2012)\citenamefont
  {Cancio~Pastor}, \citenamefont {Consolino}, \citenamefont {Giusfredi},
  \citenamefont {De~Natale}, \citenamefont {Inguscio}, \citenamefont
  {Yerokhin},\ and\ \citenamefont {Pachucki}}]{CancioPastor2012}%
  \BibitemOpen
  \bibfield  {author} {\bibinfo {author} {\bibfnamefont {P.}~\bibnamefont
  {Cancio~Pastor}}, \bibinfo {author} {\bibfnamefont {L.}~\bibnamefont
  {Consolino}}, \bibinfo {author} {\bibfnamefont {G.}~\bibnamefont
  {Giusfredi}}, \bibinfo {author} {\bibfnamefont {P.}~\bibnamefont
  {De~Natale}}, \bibinfo {author} {\bibfnamefont {M.}~\bibnamefont {Inguscio}},
  \bibinfo {author} {\bibfnamefont {V.~A.}\ \bibnamefont {Yerokhin}}, \ and\
  \bibinfo {author} {\bibfnamefont {K.}~\bibnamefont {Pachucki}},\ }\href@noop
  {} {\bibfield  {journal} {\bibinfo  {journal} {Phys. Rev. Lett.}\ }\textbf
  {\bibinfo {volume} {108}},\ \bibinfo {pages} {143001} (\bibinfo {year}
  {2012})}\BibitemShut {NoStop}%
\bibitem [{\citenamefont {Notermans}\ and\ \citenamefont
  {Vassen}(2014)}]{Notermans2014}%
  \BibitemOpen
  \bibfield  {author} {\bibinfo {author} {\bibfnamefont {R.~â. â. â.~â.}\
  \bibnamefont {Notermans}}\ and\ \bibinfo {author} {\bibfnamefont
  {W.}~\bibnamefont {Vassen}},\ }\href@noop {} {\bibfield  {journal} {\bibinfo
  {journal} {Phys. Rev. Lett.}\ }\textbf {\bibinfo {volume} {112}},\ \bibinfo
  {pages} {253002} (\bibinfo {year} {2014})}\BibitemShut {NoStop}%
\bibitem [{\citenamefont {Pohl}\ \emph {et~al.}(2010)\citenamefont {Pohl},
  \citenamefont {Antognini}, \citenamefont {Nez}, \citenamefont {Amaro},
  \citenamefont {Biraben}, \citenamefont {Cardoso}, \citenamefont {Covita},
  \citenamefont {Dax}, \citenamefont {Dhawan}, \citenamefont {Fernandes} \emph
  {et~al.}}]{Pohl2010}%
  \BibitemOpen
  \bibfield  {author} {\bibinfo {author} {\bibfnamefont {R.}~\bibnamefont
  {Pohl}}, \bibinfo {author} {\bibfnamefont {A.}~\bibnamefont {Antognini}},
  \bibinfo {author} {\bibfnamefont {F.}~\bibnamefont {Nez}}, \bibinfo {author}
  {\bibfnamefont {F.~D.}\ \bibnamefont {Amaro}}, \bibinfo {author}
  {\bibfnamefont {F.}~\bibnamefont {Biraben}}, \bibinfo {author} {\bibfnamefont
  {J.~M.}\ \bibnamefont {Cardoso}}, \bibinfo {author} {\bibfnamefont {D.~S.}\
  \bibnamefont {Covita}}, \bibinfo {author} {\bibfnamefont {A.}~\bibnamefont
  {Dax}}, \bibinfo {author} {\bibfnamefont {S.}~\bibnamefont {Dhawan}},
  \bibinfo {author} {\bibfnamefont {L.~M.}\ \bibnamefont {Fernandes}},  \emph
  {et~al.},\ }\href@noop {} {\bibfield  {journal} {\bibinfo  {journal}
  {Nature}\ }\textbf {\bibinfo {volume} {466}},\ \bibinfo {pages} {213}
  (\bibinfo {year} {2010})}\BibitemShut {NoStop}%
\bibitem [{\citenamefont {Antognini}\ \emph {et~al.}(2013)\citenamefont
  {Antognini}, \citenamefont {Nez}, \citenamefont {Schuhmann}, \citenamefont
  {Amaro}, \citenamefont {Biraben}, \citenamefont {Cardoso}, \citenamefont
  {Covita}, \citenamefont {Dax}, \citenamefont {Dhawan}, \citenamefont
  {Diepold}, \citenamefont {Fernandes}, \citenamefont {Giesen}, \citenamefont
  {Gouvea}, \citenamefont {Graf}, \citenamefont {H{\"a}nsch}, \citenamefont
  {Indelicato}, \citenamefont {Julien}, \citenamefont {Kao}, \citenamefont
  {Knowles}, \citenamefont {Kottmann}, \citenamefont {Le~Bigot}, \citenamefont
  {Liu}, \citenamefont {Lopes}, \citenamefont {Ludhova}, \citenamefont
  {Monteiro}, \citenamefont {Mulhauser}, \citenamefont {Nebel}, \citenamefont
  {Rabinowitz}, \citenamefont {dos Santos}, \citenamefont {Schaller},
  \citenamefont {Schwob}, \citenamefont {Taqqu}, \citenamefont {Veloso},
  \citenamefont {Vogelsang},\ and\ \citenamefont {Pohl}}]{Antognini2013}%
  \BibitemOpen
  \bibfield  {author} {\bibinfo {author} {\bibfnamefont {A.}~\bibnamefont
  {Antognini}}, \bibinfo {author} {\bibfnamefont {F.}~\bibnamefont {Nez}},
  \bibinfo {author} {\bibfnamefont {K.}~\bibnamefont {Schuhmann}}, \bibinfo
  {author} {\bibfnamefont {F.~D.}\ \bibnamefont {Amaro}}, \bibinfo {author}
  {\bibfnamefont {F.}~\bibnamefont {Biraben}}, \bibinfo {author} {\bibfnamefont
  {J.~M.~R.}\ \bibnamefont {Cardoso}}, \bibinfo {author} {\bibfnamefont
  {D.~S.}\ \bibnamefont {Covita}}, \bibinfo {author} {\bibfnamefont
  {A.}~\bibnamefont {Dax}}, \bibinfo {author} {\bibfnamefont {S.}~\bibnamefont
  {Dhawan}}, \bibinfo {author} {\bibfnamefont {M.}~\bibnamefont {Diepold}},
  \bibinfo {author} {\bibfnamefont {L.~M.~P.}\ \bibnamefont {Fernandes}},
  \bibinfo {author} {\bibfnamefont {A.}~\bibnamefont {Giesen}}, \bibinfo
  {author} {\bibfnamefont {A.~L.}\ \bibnamefont {Gouvea}}, \bibinfo {author}
  {\bibfnamefont {T.}~\bibnamefont {Graf}}, \bibinfo {author} {\bibfnamefont
  {T.~W.}\ \bibnamefont {H{\"a}nsch}}, \bibinfo {author} {\bibfnamefont
  {P.}~\bibnamefont {Indelicato}}, \bibinfo {author} {\bibfnamefont
  {L.}~\bibnamefont {Julien}}, \bibinfo {author} {\bibfnamefont {C.-Y.}\
  \bibnamefont {Kao}}, \bibinfo {author} {\bibfnamefont {P.}~\bibnamefont
  {Knowles}}, \bibinfo {author} {\bibfnamefont {F.}~\bibnamefont {Kottmann}},
  \bibinfo {author} {\bibfnamefont {E.-O.}\ \bibnamefont {Le~Bigot}}, \bibinfo
  {author} {\bibfnamefont {Y.-W.}\ \bibnamefont {Liu}}, \bibinfo {author}
  {\bibfnamefont {J.~A.~M.}\ \bibnamefont {Lopes}}, \bibinfo {author}
  {\bibfnamefont {L.}~\bibnamefont {Ludhova}}, \bibinfo {author} {\bibfnamefont
  {C.~M.~B.}\ \bibnamefont {Monteiro}}, \bibinfo {author} {\bibfnamefont
  {F.}~\bibnamefont {Mulhauser}}, \bibinfo {author} {\bibfnamefont
  {T.}~\bibnamefont {Nebel}}, \bibinfo {author} {\bibfnamefont
  {P.}~\bibnamefont {Rabinowitz}}, \bibinfo {author} {\bibfnamefont {J.~M.~F.}\
  \bibnamefont {dos Santos}}, \bibinfo {author} {\bibfnamefont {L.~A.}\
  \bibnamefont {Schaller}}, \bibinfo {author} {\bibfnamefont {C.}~\bibnamefont
  {Schwob}}, \bibinfo {author} {\bibfnamefont {D.}~\bibnamefont {Taqqu}},
  \bibinfo {author} {\bibfnamefont {J.~F. C.~A.}\ \bibnamefont {Veloso}},
  \bibinfo {author} {\bibfnamefont {J.}~\bibnamefont {Vogelsang}}, \ and\
  \bibinfo {author} {\bibfnamefont {R.}~\bibnamefont {Pohl}},\ }\href {\doibase
  10.1126/science.1230016} {\bibfield  {journal} {\bibinfo  {journal}
  {Science}\ }\textbf {\bibinfo {volume} {339}},\ \bibinfo {pages} {417}
  (\bibinfo {year} {2013})}\BibitemShut {NoStop}%
\bibitem [{\citenamefont {Ullmann}\ \emph {et~al.}(2017)\citenamefont
  {Ullmann}, \citenamefont {Andelkovic}, \citenamefont {Brandau}, \citenamefont
  {Dax}, \citenamefont {Geithner}, \citenamefont {Geppert}, \citenamefont
  {Gorges}, \citenamefont {Hammen}, \citenamefont {Hannen}, \citenamefont
  {Kaufmann}, \citenamefont {K\"onig}, \citenamefont {Litvinov}, \citenamefont
  {Lochmann}, \citenamefont {Maaß}, \citenamefont {Meisner}, \citenamefont
  {Murb\"ock}, \citenamefont {S\'{a}nchez}, \citenamefont {Schmidt},
  \citenamefont {Schmidt}, \citenamefont {Steck}, \citenamefont {St\"ohlker},
  \citenamefont {Thompson}, \citenamefont {Trageser}, \citenamefont
  {Vollbrecht}, \citenamefont {Weinheimer},\ and\ \citenamefont
  {N\"ortersh\"auser}}]{Ullmann2017}%
  \BibitemOpen
  \bibfield  {author} {\bibinfo {author} {\bibfnamefont {J.}~\bibnamefont
  {Ullmann}}, \bibinfo {author} {\bibfnamefont {Z.}~\bibnamefont {Andelkovic}},
  \bibinfo {author} {\bibfnamefont {C.}~\bibnamefont {Brandau}}, \bibinfo
  {author} {\bibfnamefont {A.}~\bibnamefont {Dax}}, \bibinfo {author}
  {\bibfnamefont {W.}~\bibnamefont {Geithner}}, \bibinfo {author}
  {\bibfnamefont {C.}~\bibnamefont {Geppert}}, \bibinfo {author} {\bibfnamefont
  {C.}~\bibnamefont {Gorges}}, \bibinfo {author} {\bibfnamefont
  {M.}~\bibnamefont {Hammen}}, \bibinfo {author} {\bibfnamefont
  {V.}~\bibnamefont {Hannen}}, \bibinfo {author} {\bibfnamefont
  {S.}~\bibnamefont {Kaufmann}}, \bibinfo {author} {\bibfnamefont
  {K.}~\bibnamefont {K\"onig}}, \bibinfo {author} {\bibfnamefont {Y.~A.}\
  \bibnamefont {Litvinov}}, \bibinfo {author} {\bibfnamefont {M.}~\bibnamefont
  {Lochmann}}, \bibinfo {author} {\bibfnamefont {B.}~\bibnamefont {Maaß}},
  \bibinfo {author} {\bibfnamefont {J.}~\bibnamefont {Meisner}}, \bibinfo
  {author} {\bibfnamefont {T.}~\bibnamefont {Murb\"ock}}, \bibinfo {author}
  {\bibfnamefont {R.}~\bibnamefont {S\'{a}nchez}}, \bibinfo {author}
  {\bibfnamefont {M.}~\bibnamefont {Schmidt}}, \bibinfo {author} {\bibfnamefont
  {S.}~\bibnamefont {Schmidt}}, \bibinfo {author} {\bibfnamefont
  {M.}~\bibnamefont {Steck}}, \bibinfo {author} {\bibfnamefont
  {T.}~\bibnamefont {St\"ohlker}}, \bibinfo {author} {\bibfnamefont {R.~C.}\
  \bibnamefont {Thompson}}, \bibinfo {author} {\bibfnamefont {C.}~\bibnamefont
  {Trageser}}, \bibinfo {author} {\bibfnamefont {J.}~\bibnamefont
  {Vollbrecht}}, \bibinfo {author} {\bibfnamefont {C.}~\bibnamefont
  {Weinheimer}}, \ and\ \bibinfo {author} {\bibfnamefont {W.}~\bibnamefont
  {N\"ortersh\"auser}},\ }\href@noop {} {\bibfield  {journal} {\bibinfo
  {journal} {Nat. Commun.}\ }\textbf {\bibinfo {volume} {8}},\ \bibinfo {pages}
  {15484} (\bibinfo {year} {2017})}\BibitemShut {NoStop}%
\bibitem [{\citenamefont {Yerokhin}\ and\ \citenamefont
  {Shabaev}(2015)}]{Yerokhin2015}%
  \BibitemOpen
  \bibfield  {author} {\bibinfo {author} {\bibfnamefont {V.~A.}\ \bibnamefont
  {Yerokhin}}\ and\ \bibinfo {author} {\bibfnamefont {V.~M.}\ \bibnamefont
  {Shabaev}},\ }\href {\doibase http://dx.doi.org/10.1063/1.4927487} {\bibfield
   {journal} {\bibinfo  {journal} {J. Phys. Chem. Ref. Data}\ }\textbf
  {\bibinfo {volume} {44}},\ \bibinfo {eid} {033103} (\bibinfo {year}
  {2015})}\BibitemShut {NoStop}%
\bibitem [{\citenamefont {Briand}\ \emph
  {et~al.}(1983{\natexlab{a}})\citenamefont {Briand}, \citenamefont
  {Tavernier}, \citenamefont {Indelicato}, \citenamefont {Marrus},\ and\
  \citenamefont {Gould}}]{Briand1983}%
  \BibitemOpen
  \bibfield  {author} {\bibinfo {author} {\bibfnamefont {J.~P.}\ \bibnamefont
  {Briand}}, \bibinfo {author} {\bibfnamefont {M.}~\bibnamefont {Tavernier}},
  \bibinfo {author} {\bibfnamefont {P.}~\bibnamefont {Indelicato}}, \bibinfo
  {author} {\bibfnamefont {R.}~\bibnamefont {Marrus}}, \ and\ \bibinfo {author}
  {\bibfnamefont {H.}~\bibnamefont {Gould}},\ }\href@noop {} {\bibfield
  {journal} {\bibinfo  {journal} {Phys. Rev. Lett.}\ }\textbf {\bibinfo
  {volume} {50}},\ \bibinfo {pages} {832} (\bibinfo {year}
  {1983}{\natexlab{a}})}\BibitemShut {NoStop}%
\bibitem [{\citenamefont {Briand}\ \emph
  {et~al.}(1983{\natexlab{b}})\citenamefont {Briand}, \citenamefont {Moss\'e},
  \citenamefont {Indelicato}, \citenamefont {Chevallier}, \citenamefont
  {Girard-Vernhet}, \citenamefont {Chetioui}, \citenamefont {Ramos},\ and\
  \citenamefont {Desclaux}}]{Briand1983a}%
  \BibitemOpen
  \bibfield  {author} {\bibinfo {author} {\bibfnamefont {J.~P.}\ \bibnamefont
  {Briand}}, \bibinfo {author} {\bibfnamefont {J.~P.}\ \bibnamefont {Moss\'e}},
  \bibinfo {author} {\bibfnamefont {P.}~\bibnamefont {Indelicato}}, \bibinfo
  {author} {\bibfnamefont {P.}~\bibnamefont {Chevallier}}, \bibinfo {author}
  {\bibfnamefont {D.}~\bibnamefont {Girard-Vernhet}}, \bibinfo {author}
  {\bibfnamefont {A.}~\bibnamefont {Chetioui}}, \bibinfo {author}
  {\bibfnamefont {M.~T.}\ \bibnamefont {Ramos}}, \ and\ \bibinfo {author}
  {\bibfnamefont {J.~P.}\ \bibnamefont {Desclaux}},\ }\href@noop {} {\bibfield
  {journal} {\bibinfo  {journal} {Phys. Rev. A}\ }\textbf {\bibinfo {volume}
  {28}},\ \bibinfo {pages} {1413} (\bibinfo {year}
  {1983}{\natexlab{b}})}\BibitemShut {NoStop}%
\bibitem [{\citenamefont {Briand}\ \emph {et~al.}(1984)\citenamefont {Briand},
  \citenamefont {Indelicato}, \citenamefont {Tavernier}, \citenamefont
  {Gorceix}, \citenamefont {Liesen}, \citenamefont {Beyer}, \citenamefont
  {Liu}, \citenamefont {Warczak},\ and\ \citenamefont {Desclaux}}]{Briand1984}%
  \BibitemOpen
  \bibfield  {author} {\bibinfo {author} {\bibfnamefont {J.~P.}\ \bibnamefont
  {Briand}}, \bibinfo {author} {\bibfnamefont {P.}~\bibnamefont {Indelicato}},
  \bibinfo {author} {\bibfnamefont {M.}~\bibnamefont {Tavernier}}, \bibinfo
  {author} {\bibfnamefont {O.}~\bibnamefont {Gorceix}}, \bibinfo {author}
  {\bibfnamefont {D.}~\bibnamefont {Liesen}}, \bibinfo {author} {\bibfnamefont
  {H.~F.}\ \bibnamefont {Beyer}}, \bibinfo {author} {\bibfnamefont
  {B.}~\bibnamefont {Liu}}, \bibinfo {author} {\bibfnamefont {A.}~\bibnamefont
  {Warczak}}, \ and\ \bibinfo {author} {\bibfnamefont {J.~P.}\ \bibnamefont
  {Desclaux}},\ }\href@noop {} {\bibfield  {journal} {\bibinfo  {journal} {Z.
  Physik A}\ }\textbf {\bibinfo {volume} {318}},\ \bibinfo {pages} {1}
  (\bibinfo {year} {1984})}\BibitemShut {NoStop}%
\bibitem [{\citenamefont {Richard}\ \emph {et~al.}(1984)\citenamefont
  {Richard}, \citenamefont {Stockli}, \citenamefont {Deslattes}, \citenamefont
  {Cowan}, \citenamefont {LaVilla}, \citenamefont {Johnson}, \citenamefont
  {Jones}, \citenamefont {Meron}, \citenamefont {Mann},\ and\ \citenamefont
  {Schartner}}]{Richard1984}%
  \BibitemOpen
  \bibfield  {author} {\bibinfo {author} {\bibfnamefont {P.}~\bibnamefont
  {Richard}}, \bibinfo {author} {\bibfnamefont {M.}~\bibnamefont {Stockli}},
  \bibinfo {author} {\bibfnamefont {R.}~\bibnamefont {Deslattes}}, \bibinfo
  {author} {\bibfnamefont {P.}~\bibnamefont {Cowan}}, \bibinfo {author}
  {\bibfnamefont {R.}~\bibnamefont {LaVilla}}, \bibinfo {author} {\bibfnamefont
  {B.}~\bibnamefont {Johnson}}, \bibinfo {author} {\bibfnamefont
  {K.}~\bibnamefont {Jones}}, \bibinfo {author} {\bibfnamefont
  {M.}~\bibnamefont {Meron}}, \bibinfo {author} {\bibfnamefont
  {R.}~\bibnamefont {Mann}}, \ and\ \bibinfo {author} {\bibfnamefont
  {K.}~\bibnamefont {Schartner}},\ }\href@noop {} {\bibfield  {journal}
  {\bibinfo  {journal} {Phys. Rev. A}\ }\textbf {\bibinfo {volume} {29}},\
  \bibinfo {pages} {2939} (\bibinfo {year} {1984})}\BibitemShut {NoStop}%
\bibitem [{\citenamefont {Beyer}\ \emph {et~al.}(1985)\citenamefont {Beyer},
  \citenamefont {Deslattes}, \citenamefont {Folkmann},\ and\ \citenamefont
  {LaVilla}}]{Beyer1985}%
  \BibitemOpen
  \bibfield  {author} {\bibinfo {author} {\bibfnamefont {H.~F.}\ \bibnamefont
  {Beyer}}, \bibinfo {author} {\bibfnamefont {R.~D.}\ \bibnamefont
  {Deslattes}}, \bibinfo {author} {\bibfnamefont {F.}~\bibnamefont {Folkmann}},
  \ and\ \bibinfo {author} {\bibfnamefont {R.~E.}\ \bibnamefont {LaVilla}},\
  }\href@noop {} {\bibfield  {journal} {\bibinfo  {journal} {J. Phys. B}\
  }\textbf {\bibinfo {volume} {18}},\ \bibinfo {pages} {207} (\bibinfo {year}
  {1985})}\BibitemShut {NoStop}%
\bibitem [{\citenamefont {Beyer}\ \emph {et~al.}(1991)\citenamefont {Beyer},
  \citenamefont {Indelicato}, \citenamefont {Finlayson}, \citenamefont
  {Liesen},\ and\ \citenamefont {Deslattes}}]{Beyer1991}%
  \BibitemOpen
  \bibfield  {author} {\bibinfo {author} {\bibfnamefont {H.~F.}\ \bibnamefont
  {Beyer}}, \bibinfo {author} {\bibfnamefont {P.}~\bibnamefont {Indelicato}},
  \bibinfo {author} {\bibfnamefont {K.~D.}\ \bibnamefont {Finlayson}}, \bibinfo
  {author} {\bibfnamefont {D.}~\bibnamefont {Liesen}}, \ and\ \bibinfo {author}
  {\bibfnamefont {R.~D.}\ \bibnamefont {Deslattes}},\ }\href@noop {} {\bibfield
   {journal} {\bibinfo  {journal} {Phys. Rev. A}\ }\textbf {\bibinfo {volume}
  {43}},\ \bibinfo {pages} {223} (\bibinfo {year} {1991})}\BibitemShut
  {NoStop}%
\bibitem [{\citenamefont {Briand}\ \emph {et~al.}(1990)\citenamefont {Briand},
  \citenamefont {Chevallier}, \citenamefont {Indelicato}, \citenamefont
  {Ziock},\ and\ \citenamefont {Dietrich}}]{Briand1990}%
  \BibitemOpen
  \bibfield  {author} {\bibinfo {author} {\bibfnamefont {J.~P.}\ \bibnamefont
  {Briand}}, \bibinfo {author} {\bibfnamefont {P.}~\bibnamefont {Chevallier}},
  \bibinfo {author} {\bibfnamefont {P.}~\bibnamefont {Indelicato}}, \bibinfo
  {author} {\bibfnamefont {K.~P.}\ \bibnamefont {Ziock}}, \ and\ \bibinfo
  {author} {\bibfnamefont {D.~D.}\ \bibnamefont {Dietrich}},\ }\href {\doibase
  10.1103/PhysRevLett.65.2761} {\bibfield  {journal} {\bibinfo  {journal}
  {Phys. Rev. Lett.}\ }\textbf {\bibinfo {volume} {65}},\ \bibinfo {pages}
  {2761} (\bibinfo {year} {1990})}\BibitemShut {NoStop}%
\bibitem [{\citenamefont {Lupton}\ \emph {et~al.}(1994)\citenamefont {Lupton},
  \citenamefont {Dietrich}, \citenamefont {Hailey}, \citenamefont {Stewart},\
  and\ \citenamefont {Ziock}}]{Lupton1994}%
  \BibitemOpen
  \bibfield  {author} {\bibinfo {author} {\bibfnamefont {J.~H.}\ \bibnamefont
  {Lupton}}, \bibinfo {author} {\bibfnamefont {D.~D.}\ \bibnamefont
  {Dietrich}}, \bibinfo {author} {\bibfnamefont {C.~J.}\ \bibnamefont
  {Hailey}}, \bibinfo {author} {\bibfnamefont {R.~E.}\ \bibnamefont {Stewart}},
  \ and\ \bibinfo {author} {\bibfnamefont {K.~P.}\ \bibnamefont {Ziock}},\
  }\href {\doibase 10.1103/PhysRevA.50.2150} {\bibfield  {journal} {\bibinfo
  {journal} {Phys. Rev. A}\ }\textbf {\bibinfo {volume} {50}},\ \bibinfo
  {pages} {2150} (\bibinfo {year} {1994})}\BibitemShut {NoStop}%
\bibitem [{\citenamefont {St\"ohlker}\ \emph {et~al.}(2000)\citenamefont
  {St\"ohlker}, \citenamefont {Mokler}, \citenamefont {Bosch}, \citenamefont
  {Dunford}, \citenamefont {Franzke}, \citenamefont {Klepper}, \citenamefont
  {Kozhuharov}, \citenamefont {Ludziejewski}, \citenamefont {Nolden},
  \citenamefont {Reich}, \citenamefont {Rymuza}, \citenamefont {Stachura},
  \citenamefont {Steck}, \citenamefont {Swiat},\ and\ \citenamefont
  {Warczak}}]{Stoehlker2000}%
  \BibitemOpen
  \bibfield  {author} {\bibinfo {author} {\bibfnamefont {T.}~\bibnamefont
  {St\"ohlker}}, \bibinfo {author} {\bibfnamefont {P.~H.}\ \bibnamefont
  {Mokler}}, \bibinfo {author} {\bibfnamefont {F.}~\bibnamefont {Bosch}},
  \bibinfo {author} {\bibfnamefont {R.~W.}\ \bibnamefont {Dunford}}, \bibinfo
  {author} {\bibfnamefont {F.}~\bibnamefont {Franzke}}, \bibinfo {author}
  {\bibfnamefont {O.}~\bibnamefont {Klepper}}, \bibinfo {author} {\bibfnamefont
  {C.}~\bibnamefont {Kozhuharov}}, \bibinfo {author} {\bibfnamefont
  {T.}~\bibnamefont {Ludziejewski}}, \bibinfo {author} {\bibfnamefont
  {F.}~\bibnamefont {Nolden}}, \bibinfo {author} {\bibfnamefont
  {H.}~\bibnamefont {Reich}}, \bibinfo {author} {\bibfnamefont
  {P.}~\bibnamefont {Rymuza}}, \bibinfo {author} {\bibfnamefont
  {Z.}~\bibnamefont {Stachura}}, \bibinfo {author} {\bibfnamefont
  {M.}~\bibnamefont {Steck}}, \bibinfo {author} {\bibfnamefont
  {P.}~\bibnamefont {Swiat}}, \ and\ \bibinfo {author} {\bibfnamefont
  {A.}~\bibnamefont {Warczak}},\ }\href {\doibase 10.1103/PhysRevLett.85.3109}
  {\bibfield  {journal} {\bibinfo  {journal} {Phys. Rev. Lett.}\ }\textbf
  {\bibinfo {volume} {85}},\ \bibinfo {pages} {3109} (\bibinfo {year}
  {2000})}\BibitemShut {NoStop}%
\bibitem [{\citenamefont {Gumberidze}\ \emph {et~al.}(2005)\citenamefont
  {Gumberidze}, \citenamefont {St\"ohlker}, \citenamefont
  {Bana\ifmmode~\acute{s}\else \'{s}\fi{}}, \citenamefont {Beckert},
  \citenamefont {Beller}, \citenamefont {Beyer}, \citenamefont {Bosch},
  \citenamefont {Hagmann}, \citenamefont {Kozhuharov}, \citenamefont {Liesen},
  \citenamefont {Nolden}, \citenamefont {Ma}, \citenamefont {Mokler},
  \citenamefont {Steck}, \citenamefont {Sierpowski},\ and\ \citenamefont
  {Tashenov}}]{Gumberidze2005}%
  \BibitemOpen
  \bibfield  {author} {\bibinfo {author} {\bibfnamefont {A.}~\bibnamefont
  {Gumberidze}}, \bibinfo {author} {\bibfnamefont {T.}~\bibnamefont
  {St\"ohlker}}, \bibinfo {author} {\bibfnamefont {D.}~\bibnamefont
  {Bana\ifmmode~\acute{s}\else \'{s}\fi{}}}, \bibinfo {author} {\bibfnamefont
  {K.}~\bibnamefont {Beckert}}, \bibinfo {author} {\bibfnamefont
  {P.}~\bibnamefont {Beller}}, \bibinfo {author} {\bibfnamefont {H.~F.}\
  \bibnamefont {Beyer}}, \bibinfo {author} {\bibfnamefont {F.}~\bibnamefont
  {Bosch}}, \bibinfo {author} {\bibfnamefont {S.}~\bibnamefont {Hagmann}},
  \bibinfo {author} {\bibfnamefont {C.}~\bibnamefont {Kozhuharov}}, \bibinfo
  {author} {\bibfnamefont {D.}~\bibnamefont {Liesen}}, \bibinfo {author}
  {\bibfnamefont {F.}~\bibnamefont {Nolden}}, \bibinfo {author} {\bibfnamefont
  {X.}~\bibnamefont {Ma}}, \bibinfo {author} {\bibfnamefont {P.~H.}\
  \bibnamefont {Mokler}}, \bibinfo {author} {\bibfnamefont {M.}~\bibnamefont
  {Steck}}, \bibinfo {author} {\bibfnamefont {D.}~\bibnamefont {Sierpowski}}, \
  and\ \bibinfo {author} {\bibfnamefont {S.}~\bibnamefont {Tashenov}},\ }\href
  {\doibase 10.1103/PhysRevLett.94.223001} {\bibfield  {journal} {\bibinfo
  {journal} {Phys. Rev. Lett.}\ }\textbf {\bibinfo {volume} {94}},\ \bibinfo
  {pages} {223001} (\bibinfo {year} {2005})}\BibitemShut {NoStop}%
\bibitem [{\citenamefont {Kraft-Bermuth}\ \emph {et~al.}(2017)\citenamefont
  {Kraft-Bermuth}, \citenamefont {Andrianov}, \citenamefont {Bleile},
  \citenamefont {Echler}, \citenamefont {Egelhof}, \citenamefont {Grabitz},
  \citenamefont {Ilieva}, \citenamefont {Kiselev}, \citenamefont {Kilbourne},
  \citenamefont {McCammon}, \citenamefont {Meier},\ and\ \citenamefont
  {Scholz}}]{Kraft-Bermuth2017}%
  \BibitemOpen
  \bibfield  {author} {\bibinfo {author} {\bibfnamefont {S.}~\bibnamefont
  {Kraft-Bermuth}}, \bibinfo {author} {\bibfnamefont {V.}~\bibnamefont
  {Andrianov}}, \bibinfo {author} {\bibfnamefont {A.}~\bibnamefont {Bleile}},
  \bibinfo {author} {\bibfnamefont {A.}~\bibnamefont {Echler}}, \bibinfo
  {author} {\bibfnamefont {P.}~\bibnamefont {Egelhof}}, \bibinfo {author}
  {\bibfnamefont {P.}~\bibnamefont {Grabitz}}, \bibinfo {author} {\bibfnamefont
  {S.}~\bibnamefont {Ilieva}}, \bibinfo {author} {\bibfnamefont
  {O.}~\bibnamefont {Kiselev}}, \bibinfo {author} {\bibfnamefont
  {C.}~\bibnamefont {Kilbourne}}, \bibinfo {author} {\bibfnamefont
  {D.}~\bibnamefont {McCammon}}, \bibinfo {author} {\bibfnamefont {J.~P.}\
  \bibnamefont {Meier}}, \ and\ \bibinfo {author} {\bibfnamefont
  {P.}~\bibnamefont {Scholz}},\ }\href {\doibase
  https://doi.org/10.1088/1361-6455/50/5/055603} {\bibfield  {journal}
  {\bibinfo  {journal} {J. Phys. B}\ }\textbf {\bibinfo {volume} {50}},\
  \bibinfo {pages} {055603} (\bibinfo {year} {2017})}\BibitemShut {NoStop}%
\bibitem [{\citenamefont {Hengstler}\ \emph {et~al.}(2015)\citenamefont
  {Hengstler}, \citenamefont {Keller}, \citenamefont {Sch\"otz}, \citenamefont
  {Geist}, \citenamefont {Krantz}, \citenamefont {Kempf}, \citenamefont
  {Gastaldo}, \citenamefont {Fleischmann}, \citenamefont {Gassner},
  \citenamefont {Weber}, \citenamefont {M\"artin}, \citenamefont {St\"ohlker},\
  and\ \citenamefont {Enss}}]{Hengstler2015}%
  \BibitemOpen
  \bibfield  {author} {\bibinfo {author} {\bibfnamefont {D.}~\bibnamefont
  {Hengstler}}, \bibinfo {author} {\bibfnamefont {M.}~\bibnamefont {Keller}},
  \bibinfo {author} {\bibfnamefont {C.}~\bibnamefont {Sch\"otz}}, \bibinfo
  {author} {\bibfnamefont {J.}~\bibnamefont {Geist}}, \bibinfo {author}
  {\bibfnamefont {M.}~\bibnamefont {Krantz}}, \bibinfo {author} {\bibfnamefont
  {S.}~\bibnamefont {Kempf}}, \bibinfo {author} {\bibfnamefont
  {L.}~\bibnamefont {Gastaldo}}, \bibinfo {author} {\bibfnamefont
  {A.}~\bibnamefont {Fleischmann}}, \bibinfo {author} {\bibfnamefont
  {T.}~\bibnamefont {Gassner}}, \bibinfo {author} {\bibfnamefont
  {G.}~\bibnamefont {Weber}}, \bibinfo {author} {\bibfnamefont
  {R.}~\bibnamefont {M\"artin}}, \bibinfo {author} {\bibfnamefont
  {T.}~\bibnamefont {St\"ohlker}}, \ and\ \bibinfo {author} {\bibfnamefont
  {C.}~\bibnamefont {Enss}},\ }\href
  {http://stacks.iop.org/1402-4896/2015/i=T166/a=014054} {\bibfield  {journal}
  {\bibinfo  {journal} {Phys. Scripta}\ }\textbf {\bibinfo {volume} {2015}},\
  \bibinfo {pages} {014054} (\bibinfo {year} {2015})}\BibitemShut {NoStop}%
\bibitem [{\citenamefont {Beyer}\ \emph {et~al.}(2004)\citenamefont {Beyer},
  \citenamefont {St\"ohlker}, \citenamefont {Banas}, \citenamefont {Liesen},
  \citenamefont {Protic}, \citenamefont {Beckert}, \citenamefont {Beller},
  \citenamefont {Bojowald}, \citenamefont {Bosch}, \citenamefont {Forster},
  \citenamefont {Franzke}, \citenamefont {Gumberidze}, \citenamefont {Hagmann},
  \citenamefont {Hoszowska}, \citenamefont {Indelicato}, \citenamefont
  {Klepper}, \citenamefont {Kluge}, \citenamefont {K\"onig}, \citenamefont
  {Kozhuharov}, \citenamefont {Ma}, \citenamefont {Manil}, \citenamefont
  {Mohos}, \citenamefont {Orsic-Muthig}, \citenamefont {Nolden}, \citenamefont
  {Popp}, \citenamefont {Simionovici}, \citenamefont {Sierpowski},
  \citenamefont {Spillmann}, \citenamefont {Stachura}, \citenamefont {Steck},
  \citenamefont {Tachenov}, \citenamefont {Trassinelli}, \citenamefont
  {Warczak}, \citenamefont {Wehrhan},\ and\ \citenamefont
  {Ziegler}}]{Beyer2004}%
  \BibitemOpen
  \bibfield  {author} {\bibinfo {author} {\bibfnamefont {H.}~\bibnamefont
  {Beyer}}, \bibinfo {author} {\bibfnamefont {T.}~\bibnamefont {St\"ohlker}},
  \bibinfo {author} {\bibfnamefont {D.}~\bibnamefont {Banas}}, \bibinfo
  {author} {\bibfnamefont {D.}~\bibnamefont {Liesen}}, \bibinfo {author}
  {\bibfnamefont {D.}~\bibnamefont {Protic}}, \bibinfo {author} {\bibfnamefont
  {K.}~\bibnamefont {Beckert}}, \bibinfo {author} {\bibfnamefont
  {P.}~\bibnamefont {Beller}}, \bibinfo {author} {\bibfnamefont
  {J.}~\bibnamefont {Bojowald}}, \bibinfo {author} {\bibfnamefont
  {F.}~\bibnamefont {Bosch}}, \bibinfo {author} {\bibfnamefont
  {E.}~\bibnamefont {Forster}}, \bibinfo {author} {\bibfnamefont
  {B.}~\bibnamefont {Franzke}}, \bibinfo {author} {\bibfnamefont
  {A.}~\bibnamefont {Gumberidze}}, \bibinfo {author} {\bibfnamefont
  {S.}~\bibnamefont {Hagmann}}, \bibinfo {author} {\bibfnamefont
  {J.}~\bibnamefont {Hoszowska}}, \bibinfo {author} {\bibfnamefont
  {P.}~\bibnamefont {Indelicato}}, \bibinfo {author} {\bibfnamefont
  {O.}~\bibnamefont {Klepper}}, \bibinfo {author} {\bibfnamefont {H.-J.}\
  \bibnamefont {Kluge}}, \bibinfo {author} {\bibfnamefont {S.}~\bibnamefont
  {K\"onig}}, \bibinfo {author} {\bibfnamefont {C.}~\bibnamefont {Kozhuharov}},
  \bibinfo {author} {\bibfnamefont {X.}~\bibnamefont {Ma}}, \bibinfo {author}
  {\bibfnamefont {B.}~\bibnamefont {Manil}}, \bibinfo {author} {\bibfnamefont
  {I.}~\bibnamefont {Mohos}}, \bibinfo {author} {\bibfnamefont
  {A.}~\bibnamefont {Orsic-Muthig}}, \bibinfo {author} {\bibfnamefont
  {F.}~\bibnamefont {Nolden}}, \bibinfo {author} {\bibfnamefont
  {U.}~\bibnamefont {Popp}}, \bibinfo {author} {\bibfnamefont {A.}~\bibnamefont
  {Simionovici}}, \bibinfo {author} {\bibfnamefont {D.}~\bibnamefont
  {Sierpowski}}, \bibinfo {author} {\bibfnamefont {U.}~\bibnamefont
  {Spillmann}}, \bibinfo {author} {\bibfnamefont {Z.}~\bibnamefont {Stachura}},
  \bibinfo {author} {\bibfnamefont {M.}~\bibnamefont {Steck}}, \bibinfo
  {author} {\bibfnamefont {S.}~\bibnamefont {Tachenov}}, \bibinfo {author}
  {\bibfnamefont {M.}~\bibnamefont {Trassinelli}}, \bibinfo {author}
  {\bibfnamefont {A.}~\bibnamefont {Warczak}}, \bibinfo {author} {\bibfnamefont
  {O.}~\bibnamefont {Wehrhan}}, \ and\ \bibinfo {author} {\bibfnamefont
  {E.}~\bibnamefont {Ziegler}},\ }\href@noop {} {\bibfield  {journal} {\bibinfo
   {journal} {Spectrochim. Acta Part B}\ }\textbf {\bibinfo {volume} {59}},\
  \bibinfo {pages} {1535} (\bibinfo {year} {2004})}\BibitemShut {NoStop}%
\bibitem [{\citenamefont {Chatterjee}\ \emph {et~al.}(2006)\citenamefont
  {Chatterjee}, \citenamefont {Beyer}, \citenamefont {Liesen}, \citenamefont
  {St\"ohlker}, \citenamefont {Gumberidze}, \citenamefont {Kozhuharov},
  \citenamefont {Banas}, \citenamefont {Protic}, \citenamefont {Beckert},
  \citenamefont {Beller}, \citenamefont {Krings}, \citenamefont {Bosch},
  \citenamefont {Franzke}, \citenamefont {Hagmann}, \citenamefont {Hoszowska},
  \citenamefont {Indelicato}, \citenamefont {Kluge}, \citenamefont {Ma},
  \citenamefont {Manil}, \citenamefont {Mohos}, \citenamefont {Nolden},
  \citenamefont {Popp}, \citenamefont {Simionovici}, \citenamefont
  {Sierpowski}, \citenamefont {Steck}, \citenamefont {Spillmann}, \citenamefont
  {Brandau}, \citenamefont {F\"orster}, \citenamefont {Stachura}, \citenamefont
  {Tashenov}, \citenamefont {Trassinelli}, \citenamefont {Warczak},
  \citenamefont {Wehrhan}, \citenamefont {Ziegler}, \citenamefont {Trotsenko},\
  and\ \citenamefont {Reuschl}}]{Chatterjee2006}%
  \BibitemOpen
  \bibfield  {author} {\bibinfo {author} {\bibfnamefont {S.}~\bibnamefont
  {Chatterjee}}, \bibinfo {author} {\bibfnamefont {H.~F.}\ \bibnamefont
  {Beyer}}, \bibinfo {author} {\bibfnamefont {D.}~\bibnamefont {Liesen}},
  \bibinfo {author} {\bibfnamefont {T.}~\bibnamefont {St\"ohlker}}, \bibinfo
  {author} {\bibfnamefont {A.}~\bibnamefont {Gumberidze}}, \bibinfo {author}
  {\bibfnamefont {C.}~\bibnamefont {Kozhuharov}}, \bibinfo {author}
  {\bibfnamefont {D.}~\bibnamefont {Banas}}, \bibinfo {author} {\bibfnamefont
  {D.}~\bibnamefont {Protic}}, \bibinfo {author} {\bibfnamefont
  {K.}~\bibnamefont {Beckert}}, \bibinfo {author} {\bibfnamefont
  {P.}~\bibnamefont {Beller}}, \bibinfo {author} {\bibfnamefont
  {T.}~\bibnamefont {Krings}}, \bibinfo {author} {\bibfnamefont
  {F.}~\bibnamefont {Bosch}}, \bibinfo {author} {\bibfnamefont
  {B.}~\bibnamefont {Franzke}}, \bibinfo {author} {\bibfnamefont
  {S.}~\bibnamefont {Hagmann}}, \bibinfo {author} {\bibfnamefont
  {J.}~\bibnamefont {Hoszowska}}, \bibinfo {author} {\bibfnamefont
  {P.}~\bibnamefont {Indelicato}}, \bibinfo {author} {\bibfnamefont {H.-J.}\
  \bibnamefont {Kluge}}, \bibinfo {author} {\bibfnamefont {X.}~\bibnamefont
  {Ma}}, \bibinfo {author} {\bibfnamefont {B.}~\bibnamefont {Manil}}, \bibinfo
  {author} {\bibfnamefont {I.}~\bibnamefont {Mohos}}, \bibinfo {author}
  {\bibfnamefont {F.}~\bibnamefont {Nolden}}, \bibinfo {author} {\bibfnamefont
  {U.}~\bibnamefont {Popp}}, \bibinfo {author} {\bibfnamefont {A.}~\bibnamefont
  {Simionovici}}, \bibinfo {author} {\bibfnamefont {D.}~\bibnamefont
  {Sierpowski}}, \bibinfo {author} {\bibfnamefont {M.}~\bibnamefont {Steck}},
  \bibinfo {author} {\bibfnamefont {U.}~\bibnamefont {Spillmann}}, \bibinfo
  {author} {\bibfnamefont {C.}~\bibnamefont {Brandau}}, \bibinfo {author}
  {\bibfnamefont {E.}~\bibnamefont {F\"orster}}, \bibinfo {author}
  {\bibfnamefont {Z.}~\bibnamefont {Stachura}}, \bibinfo {author}
  {\bibfnamefont {S.}~\bibnamefont {Tashenov}}, \bibinfo {author}
  {\bibfnamefont {M.}~\bibnamefont {Trassinelli}}, \bibinfo {author}
  {\bibfnamefont {A.}~\bibnamefont {Warczak}}, \bibinfo {author} {\bibfnamefont
  {O.}~\bibnamefont {Wehrhan}}, \bibinfo {author} {\bibfnamefont
  {E.}~\bibnamefont {Ziegler}}, \bibinfo {author} {\bibfnamefont
  {S.}~\bibnamefont {Trotsenko}}, \ and\ \bibinfo {author} {\bibfnamefont
  {R.}~\bibnamefont {Reuschl}},\ }\href@noop {} {\bibfield  {journal} {\bibinfo
   {journal} {Nucl. Instrum. Methods B}\ }\textbf {\bibinfo {volume} {245}},\
  \bibinfo {pages} {67} (\bibinfo {year} {2006})}\BibitemShut {NoStop}%
\bibitem [{\citenamefont {Blasche}\ and\ \citenamefont
  {Bohne}(1989)}]{Blasche1989}%
  \BibitemOpen
  \bibfield  {author} {\bibinfo {author} {\bibfnamefont {K.}~\bibnamefont
  {Blasche}}\ and\ \bibinfo {author} {\bibfnamefont {D.}~\bibnamefont
  {Bohne}},\ }in\ \href {\doibase 10.1109/PAC.1989.73027} {\emph {\bibinfo
  {booktitle} {Particle Accelerator Conference, 1989. Accelerator Science and
  Technology., Proceedings of the 1989 IEEE}}}\ (\bibinfo {year} {1989})\ pp.\
  \bibinfo {pages} {27--28 vol.1}\BibitemShut {NoStop}%
\bibitem [{\citenamefont {Beyer}\ \emph {et~al.}(2009)\citenamefont {Beyer},
  \citenamefont {Attia}, \citenamefont {Banas}, \citenamefont {Le~Bigot},
  \citenamefont {Bosch}, \citenamefont {Dousse}, \citenamefont {F\"orster},
  \citenamefont {Gumberidze}, \citenamefont {Hagmann}, \citenamefont {Hess},
  \citenamefont {Hoszowska}, \citenamefont {Indelicato}, \citenamefont
  {Jagodzinski}, \citenamefont {Kozhuharov}, \citenamefont {Krings},
  \citenamefont {Liesen}, \citenamefont {Ma}, \citenamefont {Manil},
  \citenamefont {Mohos}, \citenamefont {Pajek}, \citenamefont {Protic},
  \citenamefont {Reuschl}, \citenamefont {Rzadkiewicz}, \citenamefont
  {Simionovici}, \citenamefont {Spillmannn}, \citenamefont {Stachura},
  \citenamefont {St\"ohlker}, \citenamefont {Trassinelli}, \citenamefont
  {Trotsenko}, \citenamefont {Warczak}, \citenamefont {Wehrhan},\ and\
  \citenamefont {Ziegler}}]{Beyer2009}%
  \BibitemOpen
  \bibfield  {author} {\bibinfo {author} {\bibfnamefont {H.~F.}\ \bibnamefont
  {Beyer}}, \bibinfo {author} {\bibfnamefont {D.}~\bibnamefont {Attia}},
  \bibinfo {author} {\bibfnamefont {D.}~\bibnamefont {Banas}}, \bibinfo
  {author} {\bibfnamefont {E.-O.}\ \bibnamefont {Le~Bigot}}, \bibinfo {author}
  {\bibfnamefont {F.}~\bibnamefont {Bosch}}, \bibinfo {author} {\bibfnamefont
  {J.~C.}\ \bibnamefont {Dousse}}, \bibinfo {author} {\bibfnamefont
  {E.}~\bibnamefont {F\"orster}}, \bibinfo {author} {\bibfnamefont
  {A.}~\bibnamefont {Gumberidze}}, \bibinfo {author} {\bibfnamefont
  {S.}~\bibnamefont {Hagmann}}, \bibinfo {author} {\bibfnamefont
  {S.}~\bibnamefont {Hess}}, \bibinfo {author} {\bibfnamefont {J.}~\bibnamefont
  {Hoszowska}}, \bibinfo {author} {\bibfnamefont {P.}~\bibnamefont
  {Indelicato}}, \bibinfo {author} {\bibfnamefont {P.}~\bibnamefont
  {Jagodzinski}}, \bibinfo {author} {\bibfnamefont {C.}~\bibnamefont
  {Kozhuharov}}, \bibinfo {author} {\bibfnamefont {T.}~\bibnamefont {Krings}},
  \bibinfo {author} {\bibfnamefont {D.}~\bibnamefont {Liesen}}, \bibinfo
  {author} {\bibfnamefont {X.}~\bibnamefont {Ma}}, \bibinfo {author}
  {\bibfnamefont {B.}~\bibnamefont {Manil}}, \bibinfo {author} {\bibfnamefont
  {I.}~\bibnamefont {Mohos}}, \bibinfo {author} {\bibfnamefont
  {M.}~\bibnamefont {Pajek}}, \bibinfo {author} {\bibfnamefont
  {D.}~\bibnamefont {Protic}}, \bibinfo {author} {\bibfnamefont
  {R.}~\bibnamefont {Reuschl}}, \bibinfo {author} {\bibfnamefont
  {J.}~\bibnamefont {Rzadkiewicz}}, \bibinfo {author} {\bibfnamefont
  {A.}~\bibnamefont {Simionovici}}, \bibinfo {author} {\bibfnamefont
  {U.}~\bibnamefont {Spillmannn}}, \bibinfo {author} {\bibfnamefont
  {Z.}~\bibnamefont {Stachura}}, \bibinfo {author} {\bibfnamefont
  {T.}~\bibnamefont {St\"ohlker}}, \bibinfo {author} {\bibfnamefont
  {M.}~\bibnamefont {Trassinelli}}, \bibinfo {author} {\bibfnamefont
  {S.}~\bibnamefont {Trotsenko}}, \bibinfo {author} {\bibfnamefont
  {A.}~\bibnamefont {Warczak}}, \bibinfo {author} {\bibfnamefont
  {O.}~\bibnamefont {Wehrhan}}, \ and\ \bibinfo {author} {\bibfnamefont
  {E.}~\bibnamefont {Ziegler}},\ }\href@noop {} {\bibfield  {journal} {\bibinfo
   {journal} {Spectrochimica Acta Part B}\ }\textbf {\bibinfo {volume} {64}},\
  \bibinfo {pages} {736} (\bibinfo {year} {2009})}\BibitemShut {NoStop}%
\bibitem [{\citenamefont {Beyer}\ \emph {et~al.}(2015)\citenamefont {Beyer},
  \citenamefont {Gassner}, \citenamefont {Transsinelli}, \citenamefont {Hess},
  \citenamefont {Spillmann}, \citenamefont {Banas}, \citenamefont
  {Blumenhagen}, \citenamefont {Bosch}, \citenamefont {Brandau}, \citenamefont
  {Chen}, \citenamefont {Dimopoulou}, \citenamefont {F\"orster}, \citenamefont
  {Grisenti}, \citenamefont {Gumberidze}, \citenamefont {Hagmann},
  \citenamefont {Hillenbrand}, \citenamefont {Indelicato}, \citenamefont
  {Jagodzinski}, \citenamefont {K\"ampfer}, \citenamefont {Kozhuharov},
  \citenamefont {Lestinsky}, \citenamefont {Liesen}, \citenamefont {Litvinov},
  \citenamefont {Loetzsch}, \citenamefont {Manil}, \citenamefont {M\"artin},
  \citenamefont {Nolden}, \citenamefont {Petridis}, \citenamefont {Sanjari},
  \citenamefont {Schulze}, \citenamefont {Schwemlein}, \citenamefont
  {Simionovici}, \citenamefont {Steck}, \citenamefont {St\"ohlker},
  \citenamefont {Szabo}, \citenamefont {Trotsenko}, \citenamefont {Uschmann},
  \citenamefont {Weber}, \citenamefont {Wehrhan}, \citenamefont {Winckler},
  \citenamefont {Winters}, \citenamefont {Winters},\ and\ \citenamefont
  {Ziegler}}]{Beyer2015}%
  \BibitemOpen
  \bibfield  {author} {\bibinfo {author} {\bibfnamefont {H.~F.}\ \bibnamefont
  {Beyer}}, \bibinfo {author} {\bibfnamefont {T.}~\bibnamefont {Gassner}},
  \bibinfo {author} {\bibfnamefont {M.}~\bibnamefont {Transsinelli}}, \bibinfo
  {author} {\bibfnamefont {S.}~\bibnamefont {Hess}}, \bibinfo {author}
  {\bibfnamefont {U.}~\bibnamefont {Spillmann}}, \bibinfo {author}
  {\bibfnamefont {D.}~\bibnamefont {Banas}}, \bibinfo {author} {\bibfnamefont
  {K.-H.}\ \bibnamefont {Blumenhagen}}, \bibinfo {author} {\bibfnamefont
  {F.}~\bibnamefont {Bosch}}, \bibinfo {author} {\bibfnamefont
  {C.}~\bibnamefont {Brandau}}, \bibinfo {author} {\bibfnamefont
  {W.}~\bibnamefont {Chen}}, \bibinfo {author} {\bibfnamefont {C.}~\bibnamefont
  {Dimopoulou}}, \bibinfo {author} {\bibfnamefont {E.}~\bibnamefont
  {F\"orster}}, \bibinfo {author} {\bibfnamefont {R.~E.}\ \bibnamefont
  {Grisenti}}, \bibinfo {author} {\bibfnamefont {A.}~\bibnamefont
  {Gumberidze}}, \bibinfo {author} {\bibfnamefont {S.}~\bibnamefont {Hagmann}},
  \bibinfo {author} {\bibfnamefont {P.-M.}\ \bibnamefont {Hillenbrand}},
  \bibinfo {author} {\bibfnamefont {P.}~\bibnamefont {Indelicato}}, \bibinfo
  {author} {\bibfnamefont {P.}~\bibnamefont {Jagodzinski}}, \bibinfo {author}
  {\bibfnamefont {T.}~\bibnamefont {K\"ampfer}}, \bibinfo {author}
  {\bibfnamefont {C.}~\bibnamefont {Kozhuharov}}, \bibinfo {author}
  {\bibfnamefont {M.}~\bibnamefont {Lestinsky}}, \bibinfo {author}
  {\bibfnamefont {D.}~\bibnamefont {Liesen}}, \bibinfo {author} {\bibfnamefont
  {Y.~A.}\ \bibnamefont {Litvinov}}, \bibinfo {author} {\bibfnamefont
  {R.}~\bibnamefont {Loetzsch}}, \bibinfo {author} {\bibfnamefont
  {B.}~\bibnamefont {Manil}}, \bibinfo {author} {\bibfnamefont
  {R.}~\bibnamefont {M\"artin}}, \bibinfo {author} {\bibfnamefont
  {F.}~\bibnamefont {Nolden}}, \bibinfo {author} {\bibfnamefont
  {N.}~\bibnamefont {Petridis}}, \bibinfo {author} {\bibfnamefont {M.~S.}\
  \bibnamefont {Sanjari}}, \bibinfo {author} {\bibfnamefont {K.~S.}\
  \bibnamefont {Schulze}}, \bibinfo {author} {\bibfnamefont {M.}~\bibnamefont
  {Schwemlein}}, \bibinfo {author} {\bibfnamefont {A.}~\bibnamefont
  {Simionovici}}, \bibinfo {author} {\bibfnamefont {M.}~\bibnamefont {Steck}},
  \bibinfo {author} {\bibfnamefont {T.}~\bibnamefont {St\"ohlker}}, \bibinfo
  {author} {\bibfnamefont {C.~I.}\ \bibnamefont {Szabo}}, \bibinfo {author}
  {\bibfnamefont {S.}~\bibnamefont {Trotsenko}}, \bibinfo {author}
  {\bibfnamefont {I.}~\bibnamefont {Uschmann}}, \bibinfo {author}
  {\bibfnamefont {G.}~\bibnamefont {Weber}}, \bibinfo {author} {\bibfnamefont
  {O.}~\bibnamefont {Wehrhan}}, \bibinfo {author} {\bibfnamefont
  {N.}~\bibnamefont {Winckler}}, \bibinfo {author} {\bibfnamefont {D.~F.~A.}\
  \bibnamefont {Winters}}, \bibinfo {author} {\bibfnamefont {N.}~\bibnamefont
  {Winters}}, \ and\ \bibinfo {author} {\bibfnamefont {E.}~\bibnamefont
  {Ziegler}},\ }\href@noop {} {\bibfield  {journal} {\bibinfo  {journal} {J.
  Phys. B}\ }\textbf {\bibinfo {volume} {48}},\ \bibinfo {pages} {144010}
  (\bibinfo {year} {2015})}\BibitemShut {NoStop}%
\bibitem [{\citenamefont {B\'e}\ \emph {et~al.}(2004)\citenamefont {B\'e},
  \citenamefont {Chist\'e}, \citenamefont {Dulieu}, \citenamefont {Browne},
  \citenamefont {Chechev}, \citenamefont {Kuzmenko}, \citenamefont {Helmer},
  \citenamefont {Nichols}, \citenamefont {Sch\"onfeld},\ and\ \citenamefont
  {Dersch}}]{Be2004}%
  \BibitemOpen
  \bibfield  {author} {\bibinfo {author} {\bibfnamefont {M.-M.}\ \bibnamefont
  {B\'e}}, \bibinfo {author} {\bibfnamefont {V.}~\bibnamefont {Chist\'e}},
  \bibinfo {author} {\bibfnamefont {C.}~\bibnamefont {Dulieu}}, \bibinfo
  {author} {\bibfnamefont {E.}~\bibnamefont {Browne}}, \bibinfo {author}
  {\bibfnamefont {V.}~\bibnamefont {Chechev}}, \bibinfo {author} {\bibfnamefont
  {N.}~\bibnamefont {Kuzmenko}}, \bibinfo {author} {\bibfnamefont
  {R.}~\bibnamefont {Helmer}}, \bibinfo {author} {\bibfnamefont
  {A.}~\bibnamefont {Nichols}}, \bibinfo {author} {\bibfnamefont
  {E.}~\bibnamefont {Sch\"onfeld}}, \ and\ \bibinfo {author} {\bibfnamefont
  {R.}~\bibnamefont {Dersch}},\ }\href
  {http://www.bipm.org/utils/common/pdf/monographieRI/Monographie_BIPM-5_Tables_Vol2.pdf}
  {\emph {\bibinfo {title} {Table of Radionuclides}}},\ \bibinfo {series}
  {Monographie BIPM-5}, Vol.~\bibinfo {volume} {2}\ (\bibinfo  {publisher}
  {Bureau International des Poids et Mesures},\ \bibinfo {address} {Pavillon de
  Breteuil, F-92310 S\`evres, France},\ \bibinfo {year} {2004})\BibitemShut
  {NoStop}%
\bibitem [{\citenamefont {Spillmann}\ \emph {et~al.}(2008)\citenamefont
  {Spillmann}, \citenamefont {Br\"auning}, \citenamefont {Hess}, \citenamefont
  {Beyer}, \citenamefont {St\"ohlker}, \citenamefont {Dousse}, \citenamefont
  {Protic},\ and\ \citenamefont {Krings}}]{Spillmann2008}%
  \BibitemOpen
  \bibfield  {author} {\bibinfo {author} {\bibfnamefont {U.}~\bibnamefont
  {Spillmann}}, \bibinfo {author} {\bibfnamefont {H.}~\bibnamefont
  {Br\"auning}}, \bibinfo {author} {\bibfnamefont {S.}~\bibnamefont {Hess}},
  \bibinfo {author} {\bibfnamefont {H.}~\bibnamefont {Beyer}}, \bibinfo
  {author} {\bibfnamefont {T.}~\bibnamefont {St\"ohlker}}, \bibinfo {author}
  {\bibfnamefont {J.-C.}\ \bibnamefont {Dousse}}, \bibinfo {author}
  {\bibfnamefont {D.}~\bibnamefont {Protic}}, \ and\ \bibinfo {author}
  {\bibfnamefont {T.}~\bibnamefont {Krings}},\ }\href {\doibase
  http://dx.doi.org/10.1063/1.2963046} {\bibfield  {journal} {\bibinfo
  {journal} {Rev. Sci. Instrum.}\ }\textbf {\bibinfo {volume} {79}},\ \bibinfo
  {pages} {083101} (\bibinfo {year} {2008})}\BibitemShut {NoStop}%
\bibitem [{\citenamefont {Gassner}\ and\ \citenamefont
  {Beyer}(2015)}]{Gassner2015}%
  \BibitemOpen
  \bibfield  {author} {\bibinfo {author} {\bibfnamefont {T.}~\bibnamefont
  {Gassner}}\ and\ \bibinfo {author} {\bibfnamefont {H.~F.}\ \bibnamefont
  {Beyer}},\ }\href {http://stacks.iop.org/1402-4896/2015/i=T166/a=014052}
  {\bibfield  {journal} {\bibinfo  {journal} {Phys. Scripta}\ }\textbf
  {\bibinfo {volume} {2015}},\ \bibinfo {pages} {014052} (\bibinfo {year}
  {2015})}\BibitemShut {NoStop}%
\bibitem [{\citenamefont {Lochmann}(2013)}]{Lochmann2013}%
  \BibitemOpen
  \bibfield  {author} {\bibinfo {author} {\bibfnamefont {M.}~\bibnamefont
  {Lochmann}},\ }\emph {\bibinfo {title} {Laserspektroskopie der
  Grundzustands-Hyperfeinstruktur des lithium\"ahnlichen $^{209}Bi^{80+}$}},\
  \href@noop {} {\bibinfo {type} {{Ph.D.} thesis}},\ \bibinfo  {school}
  {Johannes Gutenberg-Universit\"at Mainz} (\bibinfo {year} {2013})\BibitemShut
  {NoStop}%
\bibitem [{\citenamefont {Lochmann}\ \emph {et~al.}(2014)\citenamefont
  {Lochmann}, \citenamefont {J\"ohren}, \citenamefont {Geppert}, \citenamefont
  {Andelkovic}, \citenamefont {Anielski}, \citenamefont {Botermann},
  \citenamefont {Bussmann}, \citenamefont {Dax}, \citenamefont {Fr\"ommgen},
  \citenamefont {Hammen}, \citenamefont {Hannen}, \citenamefont {K\"uhl},
  \citenamefont {Litvinov}, \citenamefont {L\'epez-Coto}, \citenamefont
  {St\"ohlker}, \citenamefont {Thompson}, \citenamefont {Vollbrecht},
  \citenamefont {Volotka}, \citenamefont {Weinheimer}, \citenamefont {Wen},
  \citenamefont {Will}, \citenamefont {Winters}, \citenamefont {S\'anchez},\
  and\ \citenamefont {N\"ortersh\"auser}}]{Lochmann2014}%
  \BibitemOpen
  \bibfield  {author} {\bibinfo {author} {\bibfnamefont {M.}~\bibnamefont
  {Lochmann}}, \bibinfo {author} {\bibfnamefont {R.}~\bibnamefont {J\"ohren}},
  \bibinfo {author} {\bibfnamefont {C.}~\bibnamefont {Geppert}}, \bibinfo
  {author} {\bibfnamefont {Z.}~\bibnamefont {Andelkovic}}, \bibinfo {author}
  {\bibfnamefont {D.}~\bibnamefont {Anielski}}, \bibinfo {author}
  {\bibfnamefont {B.}~\bibnamefont {Botermann}}, \bibinfo {author}
  {\bibfnamefont {M.}~\bibnamefont {Bussmann}}, \bibinfo {author}
  {\bibfnamefont {A.}~\bibnamefont {Dax}}, \bibinfo {author} {\bibfnamefont
  {N.}~\bibnamefont {Fr\"ommgen}}, \bibinfo {author} {\bibfnamefont
  {M.}~\bibnamefont {Hammen}}, \bibinfo {author} {\bibfnamefont
  {V.}~\bibnamefont {Hannen}}, \bibinfo {author} {\bibfnamefont
  {T.}~\bibnamefont {K\"uhl}}, \bibinfo {author} {\bibfnamefont {Y.~A.}\
  \bibnamefont {Litvinov}}, \bibinfo {author} {\bibfnamefont {R.}~\bibnamefont
  {L\'epez-Coto}}, \bibinfo {author} {\bibfnamefont {T.}~\bibnamefont
  {St\"ohlker}}, \bibinfo {author} {\bibfnamefont {R.~C.}\ \bibnamefont
  {Thompson}}, \bibinfo {author} {\bibfnamefont {J.}~\bibnamefont
  {Vollbrecht}}, \bibinfo {author} {\bibfnamefont {A.}~\bibnamefont {Volotka}},
  \bibinfo {author} {\bibfnamefont {C.}~\bibnamefont {Weinheimer}}, \bibinfo
  {author} {\bibfnamefont {W.}~\bibnamefont {Wen}}, \bibinfo {author}
  {\bibfnamefont {E.}~\bibnamefont {Will}}, \bibinfo {author} {\bibfnamefont
  {D.}~\bibnamefont {Winters}}, \bibinfo {author} {\bibfnamefont
  {R.}~\bibnamefont {S\'anchez}}, \ and\ \bibinfo {author} {\bibfnamefont
  {W.}~\bibnamefont {N\"ortersh\"auser}},\ }\href@noop {} {\bibfield  {journal}
  {\bibinfo  {journal} {Phys. Rev. A}\ }\textbf {\bibinfo {volume} {90}},\
  \bibinfo {pages} {030501} (\bibinfo {year} {2014})}\BibitemShut {NoStop}%
\bibitem [{\citenamefont {Brandau}(2000)}]{Brandau2000}%
  \BibitemOpen
  \bibfield  {author} {\bibinfo {author} {\bibfnamefont {C.}~\bibnamefont
  {Brandau}},\ }\emph {\bibinfo {title} {Messungen zur Photorekombination
  hochgeladener lithium\"ahnlicher}},\ \href@noop {} {\bibinfo {type} {{Ph.D.}
  thesis}},\ \bibinfo  {school} {Justus-Liebig-Universität Gie\ss en}
  (\bibinfo {year} {2000})\BibitemShut {NoStop}%
\bibitem [{\citenamefont {Beyer}\ \emph {et~al.}(1995)\citenamefont {Beyer},
  \citenamefont {Menzel}, \citenamefont {Liesen}, \citenamefont {Gallus},
  \citenamefont {Bosch}, \citenamefont {Deslattes}, \citenamefont {Indelicato},
  \citenamefont {St\"ohlker}, \citenamefont {Klepper}, \citenamefont
  {Moshammer}, \citenamefont {Nolden}, \citenamefont {Eickhoff}, \citenamefont
  {Franzke},\ and\ \citenamefont {Steck}}]{Beyer1995}%
  \BibitemOpen
  \bibfield  {author} {\bibinfo {author} {\bibfnamefont {H.}~\bibnamefont
  {Beyer}}, \bibinfo {author} {\bibfnamefont {G.}~\bibnamefont {Menzel}},
  \bibinfo {author} {\bibfnamefont {D.}~\bibnamefont {Liesen}}, \bibinfo
  {author} {\bibfnamefont {A.}~\bibnamefont {Gallus}}, \bibinfo {author}
  {\bibfnamefont {F.}~\bibnamefont {Bosch}}, \bibinfo {author} {\bibfnamefont
  {R.}~\bibnamefont {Deslattes}}, \bibinfo {author} {\bibfnamefont
  {P.}~\bibnamefont {Indelicato}}, \bibinfo {author} {\bibfnamefont
  {T.}~\bibnamefont {St\"ohlker}}, \bibinfo {author} {\bibfnamefont
  {O.}~\bibnamefont {Klepper}}, \bibinfo {author} {\bibfnamefont
  {R.}~\bibnamefont {Moshammer}}, \bibinfo {author} {\bibfnamefont
  {F.}~\bibnamefont {Nolden}}, \bibinfo {author} {\bibfnamefont
  {H.}~\bibnamefont {Eickhoff}}, \bibinfo {author} {\bibfnamefont
  {B.}~\bibnamefont {Franzke}}, \ and\ \bibinfo {author} {\bibfnamefont
  {M.}~\bibnamefont {Steck}},\ }\href {\doibase 10.1007/BF01437066} {\bibfield
  {journal} {\bibinfo  {journal} {Z. Phys. D}\ }\textbf {\bibinfo {volume}
  {35}},\ \bibinfo {pages} {169} (\bibinfo {year} {1995})}\BibitemShut
  {NoStop}%
\bibitem [{\citenamefont {St\"ohlker}\ \emph {et~al.}(2015)\citenamefont
  {St\"ohlker}, \citenamefont {Bagnoud}, \citenamefont {Blaum}, \citenamefont
  {Blazevic}, \citenamefont {Br\"auning-Demian}, \citenamefont {Durante},
  \citenamefont {Herfurth}, \citenamefont {Lestinsky}, \citenamefont
  {Litvinov}, \citenamefont {Neff}, \citenamefont {Pleskac}, \citenamefont
  {Schuch}, \citenamefont {Schippers}, \citenamefont {Severin}, \citenamefont
  {Tauschwitz}, \citenamefont {Trautmann}, \citenamefont {Varentsov},\ and\
  \citenamefont {Widmann}}]{Stoehlker2015}%
  \BibitemOpen
  \bibfield  {author} {\bibinfo {author} {\bibfnamefont {T.}~\bibnamefont
  {St\"ohlker}}, \bibinfo {author} {\bibfnamefont {V.}~\bibnamefont {Bagnoud}},
  \bibinfo {author} {\bibfnamefont {K.}~\bibnamefont {Blaum}}, \bibinfo
  {author} {\bibfnamefont {A.}~\bibnamefont {Blazevic}}, \bibinfo {author}
  {\bibfnamefont {A.}~\bibnamefont {Br\"auning-Demian}}, \bibinfo {author}
  {\bibfnamefont {M.}~\bibnamefont {Durante}}, \bibinfo {author} {\bibfnamefont
  {F.}~\bibnamefont {Herfurth}}, \bibinfo {author} {\bibfnamefont
  {M.}~\bibnamefont {Lestinsky}}, \bibinfo {author} {\bibfnamefont
  {Y.}~\bibnamefont {Litvinov}}, \bibinfo {author} {\bibfnamefont
  {S.}~\bibnamefont {Neff}}, \bibinfo {author} {\bibfnamefont {R.}~\bibnamefont
  {Pleskac}}, \bibinfo {author} {\bibfnamefont {R.}~\bibnamefont {Schuch}},
  \bibinfo {author} {\bibfnamefont {S.}~\bibnamefont {Schippers}}, \bibinfo
  {author} {\bibfnamefont {D.}~\bibnamefont {Severin}}, \bibinfo {author}
  {\bibfnamefont {A.}~\bibnamefont {Tauschwitz}}, \bibinfo {author}
  {\bibfnamefont {C.}~\bibnamefont {Trautmann}}, \bibinfo {author}
  {\bibfnamefont {D.}~\bibnamefont {Varentsov}}, \ and\ \bibinfo {author}
  {\bibfnamefont {E.}~\bibnamefont {Widmann}},\ }\href {\doibase
  http://dx.doi.org/10.1016/j.nimb.2015.07.077} {\bibfield  {journal} {\bibinfo
   {journal} {Nucl. Instrum. Methods B}\ }\textbf {\bibinfo {volume} {365}},\
  \bibinfo {pages} {680 } (\bibinfo {year} {2015})}\BibitemShut {NoStop}%
\bibitem [{\citenamefont {H\"allstr\"om}\ \emph {et~al.}(2014)\citenamefont
  {H\"allstr\"om}, \citenamefont {Bergman}, \citenamefont {Dedeo\v{g}lu},
  \citenamefont {Elg}, \citenamefont {Houtzager}, \citenamefont {Lucas},
  \citenamefont {Merev}, \citenamefont {Meisner}, \citenamefont {Schmidt},
  \citenamefont {Suomalainen},\ and\ \citenamefont {Weber}}]{Hallstrom2014}%
  \BibitemOpen
  \bibfield  {author} {\bibinfo {author} {\bibfnamefont {J.}~\bibnamefont
  {H\"allstr\"om}}, \bibinfo {author} {\bibfnamefont {A.}~\bibnamefont
  {Bergman}}, \bibinfo {author} {\bibfnamefont {S.}~\bibnamefont
  {Dedeo\v{g}lu}}, \bibinfo {author} {\bibfnamefont {A.~P.}\ \bibnamefont
  {Elg}}, \bibinfo {author} {\bibfnamefont {E.}~\bibnamefont {Houtzager}},
  \bibinfo {author} {\bibfnamefont {W.}~\bibnamefont {Lucas}}, \bibinfo
  {author} {\bibfnamefont {A.}~\bibnamefont {Merev}}, \bibinfo {author}
  {\bibfnamefont {J.}~\bibnamefont {Meisner}}, \bibinfo {author} {\bibfnamefont
  {M.}~\bibnamefont {Schmidt}}, \bibinfo {author} {\bibfnamefont {E.~P.}\
  \bibnamefont {Suomalainen}}, \ and\ \bibinfo {author} {\bibfnamefont
  {C.}~\bibnamefont {Weber}},\ }\href {\doibase 10.1109/TIM.2014.2304857}
  {\bibfield  {journal} {\bibinfo  {journal} {IEEE Trans. Instrum. Meas.}\
  }\textbf {\bibinfo {volume} {63}},\ \bibinfo {pages} {2264} (\bibinfo {year}
  {2014})}\BibitemShut {NoStop}%
\bibitem [{\citenamefont {Trassinelli}\ \emph {et~al.}(2009)\citenamefont
  {Trassinelli}, \citenamefont {Kumar}, \citenamefont {Beyer}, \citenamefont
  {Indelicato}, \citenamefont {M\"artin}, \citenamefont {Reuschl},
  \citenamefont {Kozhedub}, \citenamefont {Brandau}, \citenamefont
  {Br\"auning}, \citenamefont {Geyer}, \citenamefont {Gumberidze},
  \citenamefont {Hess}, \citenamefont {Jagodzinski}, \citenamefont
  {Kozhuharov}, \citenamefont {Liesen}, \citenamefont {Spillmann},
  \citenamefont {Trotsenko}, \citenamefont {Weber}, \citenamefont {Winters},\
  and\ \citenamefont {St\"ohlker}}]{Trassinelli2009}%
  \BibitemOpen
  \bibfield  {author} {\bibinfo {author} {\bibfnamefont {M.}~\bibnamefont
  {Trassinelli}}, \bibinfo {author} {\bibfnamefont {A.}~\bibnamefont {Kumar}},
  \bibinfo {author} {\bibfnamefont {H.}~\bibnamefont {Beyer}}, \bibinfo
  {author} {\bibfnamefont {P.}~\bibnamefont {Indelicato}}, \bibinfo {author}
  {\bibfnamefont {R.}~\bibnamefont {M\"artin}}, \bibinfo {author}
  {\bibfnamefont {R.}~\bibnamefont {Reuschl}}, \bibinfo {author} {\bibfnamefont
  {Y.}~\bibnamefont {Kozhedub}}, \bibinfo {author} {\bibfnamefont
  {C.}~\bibnamefont {Brandau}}, \bibinfo {author} {\bibfnamefont
  {H.}~\bibnamefont {Br\"auning}}, \bibinfo {author} {\bibfnamefont
  {S.}~\bibnamefont {Geyer}}, \bibinfo {author} {\bibfnamefont
  {A.}~\bibnamefont {Gumberidze}}, \bibinfo {author} {\bibfnamefont
  {S.}~\bibnamefont {Hess}}, \bibinfo {author} {\bibfnamefont {P.}~\bibnamefont
  {Jagodzinski}}, \bibinfo {author} {\bibfnamefont {C.}~\bibnamefont
  {Kozhuharov}}, \bibinfo {author} {\bibfnamefont {D.}~\bibnamefont {Liesen}},
  \bibinfo {author} {\bibfnamefont {U.}~\bibnamefont {Spillmann}}, \bibinfo
  {author} {\bibfnamefont {S.}~\bibnamefont {Trotsenko}}, \bibinfo {author}
  {\bibfnamefont {G.}~\bibnamefont {Weber}}, \bibinfo {author} {\bibfnamefont
  {D.}~\bibnamefont {Winters}}, \ and\ \bibinfo {author} {\bibfnamefont
  {T.}~\bibnamefont {St\"ohlker}},\ }\href {\doibase
  https://doi.org/10.1209/0295-5075/87/63001} {\bibfield  {journal} {\bibinfo
  {journal} {Eur. Phys. Lett.}\ }\textbf {\bibinfo {volume} {87}},\ \bibinfo
  {pages} {63001} (\bibinfo {year} {2009})}\BibitemShut {NoStop}%
\bibitem [{\citenamefont {Chantler}\ \emph {et~al.}(2012)\citenamefont
  {Chantler}, \citenamefont {Kinnane}, \citenamefont {Gillaspy}, \citenamefont
  {Hudson}, \citenamefont {Payne}, \citenamefont {Smale}, \citenamefont
  {Henins}, \citenamefont {Pomeroy}, \citenamefont {Tan}, \citenamefont
  {Kimpton}, \citenamefont {Takacs},\ and\ \citenamefont
  {Makonyi}}]{Chantler2012}%
  \BibitemOpen
  \bibfield  {author} {\bibinfo {author} {\bibfnamefont {C.~T.}\ \bibnamefont
  {Chantler}}, \bibinfo {author} {\bibfnamefont {M.~N.}\ \bibnamefont
  {Kinnane}}, \bibinfo {author} {\bibfnamefont {J.~D.}\ \bibnamefont
  {Gillaspy}}, \bibinfo {author} {\bibfnamefont {L.~T.}\ \bibnamefont
  {Hudson}}, \bibinfo {author} {\bibfnamefont {A.~T.}\ \bibnamefont {Payne}},
  \bibinfo {author} {\bibfnamefont {L.~F.}\ \bibnamefont {Smale}}, \bibinfo
  {author} {\bibfnamefont {A.}~\bibnamefont {Henins}}, \bibinfo {author}
  {\bibfnamefont {J.~M.}\ \bibnamefont {Pomeroy}}, \bibinfo {author}
  {\bibfnamefont {J.~N.}\ \bibnamefont {Tan}}, \bibinfo {author} {\bibfnamefont
  {J.~A.}\ \bibnamefont {Kimpton}}, \bibinfo {author} {\bibfnamefont
  {E.}~\bibnamefont {Takacs}}, \ and\ \bibinfo {author} {\bibfnamefont
  {K.}~\bibnamefont {Makonyi}},\ }\href {\doibase
  http://dx.doi.org/10.1103/PhysRevLett.109.153001} {\bibfield  {journal}
  {\bibinfo  {journal} {Phys. Rev. Lett.}\ }\textbf {\bibinfo {volume} {109}},\
  \bibinfo {pages} {153001} (\bibinfo {year} {2012})}\BibitemShut {NoStop}%
\bibitem [{\citenamefont {Amaro}\ \emph {et~al.}(2012)\citenamefont {Amaro},
  \citenamefont {Schlesser}, \citenamefont {Guerra}, \citenamefont {Bigot},
  \citenamefont {Isac}, \citenamefont {Travers}, \citenamefont {Santos},
  \citenamefont {Szabo}, \citenamefont {Gumberidze},\ and\ \citenamefont
  {Indelicato}}]{Amaro2012}%
  \BibitemOpen
  \bibfield  {author} {\bibinfo {author} {\bibfnamefont {P.}~\bibnamefont
  {Amaro}}, \bibinfo {author} {\bibfnamefont {S.}~\bibnamefont {Schlesser}},
  \bibinfo {author} {\bibfnamefont {M.}~\bibnamefont {Guerra}}, \bibinfo
  {author} {\bibfnamefont {E.-O.~L.}\ \bibnamefont {Bigot}}, \bibinfo {author}
  {\bibfnamefont {J.-M.}\ \bibnamefont {Isac}}, \bibinfo {author}
  {\bibfnamefont {P.}~\bibnamefont {Travers}}, \bibinfo {author} {\bibfnamefont
  {J.~P.}\ \bibnamefont {Santos}}, \bibinfo {author} {\bibfnamefont {C.~I.}\
  \bibnamefont {Szabo}}, \bibinfo {author} {\bibfnamefont {A.}~\bibnamefont
  {Gumberidze}}, \ and\ \bibinfo {author} {\bibfnamefont {P.}~\bibnamefont
  {Indelicato}},\ }\href@noop {} {\bibfield  {journal} {\bibinfo  {journal}
  {Phys. Rev. Lett.}\ }\textbf {\bibinfo {volume} {109}},\ \bibinfo {pages}
  {043005} (\bibinfo {year} {2012})}\BibitemShut {NoStop}%
\bibitem [{\citenamefont {Rudolph}\ \emph {et~al.}(2013)\citenamefont
  {Rudolph}, \citenamefont {Bernitt}, \citenamefont {Epp}, \citenamefont
  {Steinbr\"ugge}, \citenamefont {Beilmann}, \citenamefont {Brown},
  \citenamefont {Eberle}, \citenamefont {Graf}, \citenamefont {Harman},
  \citenamefont {Hell}, \citenamefont {Leutenegger}, \citenamefont {M\"uller},
  \citenamefont {Schlage}, \citenamefont {Wille}, \citenamefont {Yava\c{s}},
  \citenamefont {Ullrich},\ and\ \citenamefont
  {L\'opez-Urrutia}}]{Rudolph2013}%
  \BibitemOpen
  \bibfield  {author} {\bibinfo {author} {\bibfnamefont {J.~K.}\ \bibnamefont
  {Rudolph}}, \bibinfo {author} {\bibfnamefont {S.}~\bibnamefont {Bernitt}},
  \bibinfo {author} {\bibfnamefont {S.~W.}\ \bibnamefont {Epp}}, \bibinfo
  {author} {\bibfnamefont {R.}~\bibnamefont {Steinbr\"ugge}}, \bibinfo {author}
  {\bibfnamefont {C.}~\bibnamefont {Beilmann}}, \bibinfo {author}
  {\bibfnamefont {G.~V.}\ \bibnamefont {Brown}}, \bibinfo {author}
  {\bibfnamefont {S.}~\bibnamefont {Eberle}}, \bibinfo {author} {\bibfnamefont
  {A.}~\bibnamefont {Graf}}, \bibinfo {author} {\bibfnamefont {Z.}~\bibnamefont
  {Harman}}, \bibinfo {author} {\bibfnamefont {N.}~\bibnamefont {Hell}},
  \bibinfo {author} {\bibfnamefont {M.}~\bibnamefont {Leutenegger}}, \bibinfo
  {author} {\bibfnamefont {A.}~\bibnamefont {M\"uller}}, \bibinfo {author}
  {\bibfnamefont {K.}~\bibnamefont {Schlage}}, \bibinfo {author} {\bibfnamefont
  {H.~C.}\ \bibnamefont {Wille}}, \bibinfo {author} {\bibfnamefont
  {H.}~\bibnamefont {Yava\c{s}}}, \bibinfo {author} {\bibfnamefont
  {J.}~\bibnamefont {Ullrich}}, \ and\ \bibinfo {author} {\bibfnamefont
  {J.~R.~C.}\ \bibnamefont {L\'opez-Urrutia}},\ }\href@noop {} {\bibfield
  {journal} {\bibinfo  {journal} {Phys. Rev. Lett.}\ }\textbf {\bibinfo
  {volume} {111}},\ \bibinfo {pages} {103002} (\bibinfo {year}
  {2013})}\BibitemShut {NoStop}%
\bibitem [{\citenamefont {Kubi\v{c}ek}\ \emph {et~al.}(2014)\citenamefont
  {Kubi\v{c}ek}, \citenamefont {Mokler}, \citenamefont {M\"ackel},
  \citenamefont {Ullrich},\ and\ \citenamefont
  {L\'opez-Urrutia}}]{Kubicek2014}%
  \BibitemOpen
  \bibfield  {author} {\bibinfo {author} {\bibfnamefont {K.}~\bibnamefont
  {Kubi\v{c}ek}}, \bibinfo {author} {\bibfnamefont {P.~H.}\ \bibnamefont
  {Mokler}}, \bibinfo {author} {\bibfnamefont {V.}~\bibnamefont {M\"ackel}},
  \bibinfo {author} {\bibfnamefont {J.}~\bibnamefont {Ullrich}}, \ and\
  \bibinfo {author} {\bibfnamefont {J.~R.~C.}\ \bibnamefont
  {L\'opez-Urrutia}},\ }\href@noop {} {\bibfield  {journal} {\bibinfo
  {journal} {Phys. Rev. A}\ }\textbf {\bibinfo {volume} {90}},\ \bibinfo
  {pages} {032508} (\bibinfo {year} {2014})}\BibitemShut {NoStop}%
\bibitem [{\citenamefont {Payne}\ \emph {et~al.}(2014)\citenamefont {Payne},
  \citenamefont {Chantler}, \citenamefont {Kinnane}, \citenamefont {Gillaspy},
  \citenamefont {Hudson}, \citenamefont {Smale}, \citenamefont {Henins},
  \citenamefont {Kimpton},\ and\ \citenamefont {Takacs}}]{Payne2014}%
  \BibitemOpen
  \bibfield  {author} {\bibinfo {author} {\bibfnamefont {A.~T.}\ \bibnamefont
  {Payne}}, \bibinfo {author} {\bibfnamefont {C.~T.}\ \bibnamefont {Chantler}},
  \bibinfo {author} {\bibfnamefont {M.~N.}\ \bibnamefont {Kinnane}}, \bibinfo
  {author} {\bibfnamefont {J.~D.}\ \bibnamefont {Gillaspy}}, \bibinfo {author}
  {\bibfnamefont {L.~T.}\ \bibnamefont {Hudson}}, \bibinfo {author}
  {\bibfnamefont {L.~F.}\ \bibnamefont {Smale}}, \bibinfo {author}
  {\bibfnamefont {A.}~\bibnamefont {Henins}}, \bibinfo {author} {\bibfnamefont
  {J.~A.}\ \bibnamefont {Kimpton}}, \ and\ \bibinfo {author} {\bibfnamefont
  {E.}~\bibnamefont {Takacs}},\ }\href {\doibase
  https://doi.org/10.1088/0953-4075/47/18/185001} {\bibfield  {journal}
  {\bibinfo  {journal} {J. Phys. B}\ }\textbf {\bibinfo {volume} {47}},\
  \bibinfo {pages} {185001} (\bibinfo {year} {2014})}\BibitemShut {NoStop}%
\bibitem [{\citenamefont {Epp}\ \emph {et~al.}(2015)\citenamefont {Epp},
  \citenamefont {Steinbr\"ugge}, \citenamefont {Bernitt}, \citenamefont
  {Rudolph}, \citenamefont {Beilmann}, \citenamefont {Bekker}, \citenamefont
  {M\"uller}, \citenamefont {Versolato}, \citenamefont {Wille}, \citenamefont
  {Yavas}, \citenamefont {Ullrich},\ and\ \citenamefont
  {L$\acute{o}$pez-Urrutia}}]{Epp2015}%
  \BibitemOpen
  \bibfield  {author} {\bibinfo {author} {\bibfnamefont {S.~W.}\ \bibnamefont
  {Epp}}, \bibinfo {author} {\bibfnamefont {R.}~\bibnamefont {Steinbr\"ugge}},
  \bibinfo {author} {\bibfnamefont {S.}~\bibnamefont {Bernitt}}, \bibinfo
  {author} {\bibfnamefont {J.~K.}\ \bibnamefont {Rudolph}}, \bibinfo {author}
  {\bibfnamefont {C.}~\bibnamefont {Beilmann}}, \bibinfo {author}
  {\bibfnamefont {H.}~\bibnamefont {Bekker}}, \bibinfo {author} {\bibfnamefont
  {A.}~\bibnamefont {M\"uller}}, \bibinfo {author} {\bibfnamefont {O.~O.}\
  \bibnamefont {Versolato}}, \bibinfo {author} {\bibfnamefont {H.-C.}\
  \bibnamefont {Wille}}, \bibinfo {author} {\bibfnamefont {H.}~\bibnamefont
  {Yavas}}, \bibinfo {author} {\bibfnamefont {J.}~\bibnamefont {Ullrich}}, \
  and\ \bibinfo {author} {\bibfnamefont {J.~R.~C.}\ \bibnamefont
  {L$\acute{o}$pez-Urrutia}},\ }\href {\doibase
  https://doi.org/10.1103/PhysRevA.92.020502} {\bibfield  {journal} {\bibinfo
  {journal} {Phys. Rev. A}\ }\textbf {\bibinfo {volume} {92}},\ \bibinfo
  {pages} {020502(R)} (\bibinfo {year} {2015})}\BibitemShut {NoStop}%
\bibitem [{\citenamefont {Beiersdorfer}\ and\ \citenamefont
  {Brown}(2015)}]{Beiersdorfer2015}%
  \BibitemOpen
  \bibfield  {author} {\bibinfo {author} {\bibfnamefont {P.}~\bibnamefont
  {Beiersdorfer}}\ and\ \bibinfo {author} {\bibfnamefont {G.~V.}\ \bibnamefont
  {Brown}},\ }\href {\doibase https://doi.org/10.1103/PhysRevA.91.032514}
  {\bibfield  {journal} {\bibinfo  {journal} {Phys. Rev. A}\ }\textbf {\bibinfo
  {volume} {91}},\ \bibinfo {pages} {032514} (\bibinfo {year}
  {2015})}\BibitemShut {NoStop}%
\end{thebibliography}%

\end{document}